\documentclass[aps,showkeys,twocolumn,preprintnumbers,amsmath,amssymb,superscriptaddress,floatfix,nofootinbib]{revtex4}
\usepackage{graphicx,color,dcolumn,booktabs,bm}
\usepackage{amsmath}
\usepackage{graphicx}
\usepackage{longtable,lscape}
\usepackage{txfonts}
\usepackage{overpic}
\usepackage{epsfig}
\usepackage{amssymb}
\usepackage{rotating}
\usepackage{epstopdf}
\usepackage{appendix}
\usepackage{indentfirst}
\usepackage{feynmf}   
\usepackage{slashed}  
\usepackage{cases}
\usepackage{color}
\usepackage{multirow}
\usepackage{graphicx,color,dcolumn,booktabs,bm}
\usepackage{cases}
\usepackage{array}

\graphicspath{{Figures/}} %

\usepackage[colorlinks, citecolor=blue,anchorcolor=red,menucolor=red, linkcolor=red,filecolor=red,runcolor=red,urlcolor=blue,frenchlinks=red]{hyperref}

\begin{document}
\title{Bottom-charmed baryons in a nonrelativistic quark model}
\author{Qing-Fu Song}

\affiliation{Department of Physics, Hunan Normal University, and Key Laboratory of Low-Dimensional
Quantum Structures and Quantum Control of Ministry of Education, Changsha 410081, China}
\affiliation{Key Laboratory for Matter Microstructure and Function of Hunan Province, Hunan Normal University, Changsha 410081, China}

\author{Qi-Fang L\"{u}}\email{lvqifang@hunnu.edu.cn}

\affiliation{Department of Physics, Hunan Normal University, and Key Laboratory of Low-Dimensional
Quantum Structures and Quantum Control of Ministry of Education, Changsha 410081, China}
\affiliation{Key Laboratory for Matter Microstructure and Function of Hunan Province, Hunan Normal University, Changsha 410081, China}

\affiliation{Research Center for Nuclear Physics (RCNP), Ibaraki, Osaka 567-0047, Japan}

\author{Atsushi Hosaka}\email{hosaka@rcnp.osaka-u.ac.jp
} %
\affiliation{Research Center for Nuclear Physics (RCNP), Ibaraki, Osaka 567-0047, Japan}

\affiliation{Advanced Science Research Center, Japan Atomic Energy Agency, Tokai, Ibaraki 319-1195, Japan}

\begin{abstract}
In this work, we study the low-lying mass spectra for bottom-charmed baryons in a nonrelativistic quark model by solving the three-body  Schr\"odinger equation. The lowest $\Xi_{bc}$, $\Xi_{bc}^\prime$, $\Omega_{bc}$, and $\Omega_{bc}^\prime$ states are predicted to be about 6979, 6953, 7109, and 7092 MeV, respectively. Also, the strong decays for the low-lying excited states are investigated. Our results indicate that some of $\lambda-$mode $P-$wave bottom-charmed baryons are relatively narrow, which can be searched for  in future  experiments. For the low-lying $\rho-$mode and $\rho-\lambda$ hybrid states, their strong decays  are highly suppressed and they can survive as extremely  narrow states. Moreover, the mass spectra and strong decays for bottom-charmed baryons preserve the heavy quark symmetry well. We hope our calculations can provide helpful information for further experimental and theoretical researches.

\end{abstract}


\maketitle

\section{Introduction}
In the past years, with the development of the large-scale accelerator facilities, plenty of heavy baryons have been observed and significant progress has been achieved in experiments. These discoveries have triggered wide attentions of theorists, which leads the study on mass spectra and internal structures of heavy baryons to a hot topic in hadron physics~\cite{Chen:2016spr,Cheng:2021qpd}. Understanding the nature of heavy baryons and searching for the missing heavy resonances can help us to establish and complete the hadron spectroscopy and provide a good platform to better investigate the heavy quark symmetry. Until now, most experimental observations in heavy baryon sector belong to the singly heavy baryons, while the doubly heavy baryons are rare. Due to the lack of experimental data, our understanding of the doubly heavy baryons is still scarce, and more experimental and theoretical efforts are encouraged and needed.

In 2002, the SELEX Collaboration reported an evidence of a doubly charmed baryon $\Xi_{cc}^{+}$ that has a mass of 3519 $\pm$ 1 MeV in the $\Lambda_c^+K^-\pi^+$ final state~\cite{Mattson:2002vu} and $pD^+K^-$ decay mode~\cite{Ocherashvili:2004hi}. However, the existence of $\Xi_{cc}^{+}(3519)$ was disfavored by the following FOCUS, BaBar, Belle and LHCb Collaborations~\cite{Ratti:2003ez,Aubert:2006qw,Chistov:2006zj,Aaij:2013voa}, and the following theoretical works do not also support this discovery.  In 2017, a highly significant structure $\Xi_{cc}^{++}(3621)$ with a mass of 3621.40 $\pm$ 0.72 $\pm$ 0.27 $\pm$ 0.14 MeV was observed in the $\Lambda_{c}^{+}K^{-}\pi^{+}\pi^{+}$ mass spectrum by the LHCb Collaboration~\cite{Aaij:2017ueg}. Subsequently, the LHCb Collaboration measured its lifetime~\cite{LHCb:2018zpl,LHCb:2019qed} and other decay modes~\cite{LHCb:2018pcs,LHCb:2019ybf,LHCb:2019epo,LHCb:2022rpd}. Meanwhile, the LHCb Collaboration paid lots of attentions to hunt for more doubly heavy baryons, however no more signal has been found so far~\cite{LHCb:2021eaf, LHCb:2020iko, LHCb:2021xba, LHCb:2021rkb,LHCb:2019gqy,LHCb:2022fbu}. In particular, the $\Xi_{bc}$ and $\Xi_{bc}^{\prime}$ state was placed great expectations to be discovered in the near future, where one charm quark in $\Xi_{cc}$ is replaced by a bottom quark.

Theoretically, there are various methods to predict the mass spectra of doubly heavy baryons, such as potential models~\cite{Kiselev:2001fw,Ebert:1996ec,Tong:1999qs,Ebert:2002ig,Gershtein:2000nx,Roberts:2007ni,Giannuzzi:2009gh,Martynenko:2007je,Valcarce:2008dr,Eakins:2012jk,Shah:2017liu,Wang:2021rjk,Yu:2022lel,Li:2022ywz,Soto:2020pfa},  heavy quark symmetry and mass formulas~\cite{Savage:1990di,Song:2022csw,Roncaglia:1995az,Cohen:2006jg,Karliner:2014gca,Wei:2015gsa,Wei:2016jyk,Oudichhya:2022ssc}, QCD sum rule~\cite{Zhang:2008rt,Tang:2011fv,Wang:2010hs,Aliev:2012ru,Aliev:2012nn,Aliev:2012iv}, lattice QCD~\cite{Liu:2009jc,Brown:2014ena,Padmanath:2015jea,Mathur:2018rwu,Mathur:2018epb}, and so on. Besides the mass spectra, the weak and radiative decays of the doubly heavy baryons are also widely discussed in the literature~\cite{Faessler:2001mr,Faessler:2009xn,Albertus:2009ww,White:1991hz,Li:2017ndo,Yu:2017zst,Ebert:2004ck,Roberts:2008wq,Branz:2010pq,Hackman:1977am,Bernotas:2013eia,Dai:2000hza,Albertus:2010hi,Qin:2021zqx,Bahtiyar:2018vub}, which provide helpful information for the experimental searches. Among these  fruitful theoretical studies, there were only a few works on strong decay behaviors~\cite{Eakins:2012fq,Xiao:2017udy,Mehen:2017nrh,Ma:2017nik,Xiao:2017dly,Yan:2018zdt,He:2021iwx,Chen:2022fye}. Based on the experimental and theoretical status, the strong decay of doubly heavy baryons should be urgently investigated and  highly valued. 

In this work, we concentrate on the bottom-charmed family that is made up of a bottom quark $b$, a charm quark $c$ and a light quark ($u$, $d$, or $s$). When the light quark belongs to up or down quark, the bottom-charmed baryon is named as $\Xi_{bc}$ or $\Xi_{bc}^{\prime}$; When the light quark is a strange quark, the bottom-charmed baryon is denoted as $\Omega_{bc}$ or $\Omega_{bc}^{\prime}$. The study of bottom-charmed baryons does not only provide an opportunity for us to investigate the heavy quark symmetry and chiral dynamics simultaneously, but also supplies a unique platform about the conventional baryons with three non-identical quarks. Moreover, the excited states and their strong decay behaviors are essential for  better understanding the spectroscopy for bottom-charmed baryons and helping experimentalists to hunt for more hadrons. Among various properties, the Okubo-Zweig-Iizuka-allowed (OZI-allowed) two-body decay processes are particularly interesting where one light meson is, especially the pion, emitted from the parent baryon. 
In this process, the meson couples to the light quark, and the heavy quark subsystem behaves simply as a spectator. Hence this provides a good platform to investigate dynamics of chiral symmetry at the single quark level.  However, the study on strong decays for bottom-charmed baryon is very scarce~\cite{Eakins:2012fq}, and then it is time to explore this topic systematically.

In a previous work, a nonrelativistic quark model was adopted to study the mass spectrum of heavy baryons with two identical quarks, and gained significant achievements~\cite{Yoshida:2015tia}. However, the case of all three quarks with different masses were not studied. Here, we employ the same quark model to bottom-charmed system by solving the three-body Schr\"odinger equation in order to get the mass spectrum consistently. Besides the masses, the realistic wave functions are obtained simultaneously, and can be used for strong decay calculations. The lowest $\Xi_{bc}$, $\Xi_{bc}^\prime$, $\Omega_{bc}$, and $\Omega_{bc}^\prime$ states are predicted to be about 6979, 6953, 7109, and 7092 MeV, respectively. Our results indicate that some of $\lambda-$mode $\Xi_{bc}(1P)$, $\Xi_{bc}^{\prime}(1P)$, $\Omega_{bc}(1P)$, $\Omega_{bc}^{\prime}(1P)$ states are narrow, which have good potentials to be observed by future  experiments. Also, the strong decays of the low-lying $\rho-$mode and $\rho- \lambda$ hybrid states are highly suppressed and  can be searched for in the processes with  electromagnetic and weak interactions.

This paper is organized as follows. The formalism of potential model and pseudoscalar meson emissions is briefly introduced in Sec.~\ref{FORMALISM}. We present the numerical results and discussions for the bottom-charmed baryons in Sec.~\ref{low-lying}. A summary is given in the last section.

\section{FORMALISM}{\label{FORMALISM}}
\subsection{Potential Model}
 In order to calculate the spectrum of the low-lying bottom-charmed baryons, we adopt a nonrelativistic quark  model, where the Hamiltonian can be expressed as
\begin{equation}
H=T+\sum_{i<j}^{3}V(r_{ij})\label{h}
\end{equation}
with the kinetic energy
 \begin{equation}
    T= \sum_{i=1}^{3}\left( m_{i}+\frac{\boldsymbol{p_{i}^{2}}}{2m_{i}}\right)-T_{CM}
\end{equation}
and the effective potential 
\begin{equation}
V\left(r_{ij}\right)=V^{conf}\left(r_{ij}\right)+V^{coul}\left(r_{ij}\right)+V^{SD}\left(r_{ij}\right).
\end{equation}
Here $T_{CM}$ is the center-of-mass energy, $r_{ij}$ is the distance between the $i$th  and $j$th quarks. The linear confinement potential $V^{conf}\left(r_{ij}\right)$ and one-gluon-exchange potential $V^{coul}\left(r_{ij}\right)$ are as follows
\begin{equation}
V^{conf}\left(r_{ij}\right)=\frac{br_{ij}}{2}+C,
\end{equation}
\begin{equation}
     V^{coul}\left(r_{ij}\right)=-\frac{2\alpha^{coul}}{3r_{ij}},
\end{equation}
where $C$ is the overall zero-point-energy parameter. The spin-dependent interaction $V^{SD}\left(r_{ij}\right)$ is the sum of spin-spin term $V^{SS}\left(r_{ij}\right)$, spin-orbit term $V^{LS}\left(r_{ij}\right)$, and tensor term $V^{Ten}\left(r_{ij}\right)$ 
\begin{equation}
V^{SS}\left({r_{ij}}\right)=\frac{16\pi \alpha^{ss}}{9m_{i}m_{j}}\boldsymbol{s_{i}}\cdot\boldsymbol{s_{j}}\frac{\Lambda^{2}}{4\pi r_{ij}}\exp(-\Lambda r_{ij}),
\end{equation}

\begin{eqnarray}
V^{LS}\left({r_{ij}}\right)&=&\frac{{\alpha}^{\rm{so}}(1-\exp(-\Lambda r_{ij}))^2}{3{r_{ij}}^3} \nonumber \\
&&\times\bigg[\bigg(\frac{1}{m_i^2}+\frac{1}{m_j^2}+\frac{4}{m_im_j}\bigg)\bm{L}_{ij}\cdot\left(\bm{s}_i+\bm{s}_j\right)\nonumber\\
&&+\bigg(\frac{1}{m_i^2}-\frac{1}{m_j^2}\bigg)\bm{L}_{ij}\cdot\left(\bm{s}_i-\bm{s}_j\right)\bigg],
\end{eqnarray}

\begin{eqnarray}
V^{Ten}\left({r_{ij}}\right)&=&\left. \frac{{2\alpha}^{\rm{ten}}(1-\exp(-\Lambda r_{ij}))^2}{3m_im_j{r_{ij}}^3}\left(\frac{3(\bm{s}_i\cdot\bm{ r_{ij}})(\bm{s}_j\cdot\bm{ r_{ij}})}{{r_{ij}}^2}\right.\right.\nonumber\\
&&-\left.\bm{s}_i\cdot\bm{s}_j\biggr)\right.. \label{hyp}
\end{eqnarray}
The  parameters are taken from the original work~\cite{Yoshida:2015tia} and listed in the Table~\ref{Quark model parameters}. Also, the overall constant $C$ is adjusted by fixing the mass of ground state $\Xi_{cc}(3621)$, which is suitable for present bottom-charmed baryons systems.

\begin{table}[htbp]
\caption{\label{Quark model parameters} The relevant parameters adopted in this work.}
\begin{ruledtabular}
\begin{tabular}{ccccccccc}
&Parameters
&Value
\\\hline
&$m_{u/d}(\textrm{GeV})$&0.300\\
&$m_s$~$(\textrm{GeV})$&0.510\\
&$m_c$$(\textrm{GeV})$ &1.750\\
&$m_b$$(\textrm{GeV})$&5.112\\
&$b$ $(\textrm{GeV}^{2})$&0.165\\
&$K$ $(\textrm{GeV})$&0.090\\
&$\alpha^{ss}$&1.200\\
&$\alpha^{so}$&0.077\\
&$\alpha^{ten}$&0.077\\
&$\Lambda$$(\textrm{fm}^{-1})$&3.500\\
&${C}$$(\textrm{GeV})$&-1.203\\
\end{tabular}
\end{ruledtabular}
\end{table}

\begin{figure}[! htpb]
	\centering
	\includegraphics[scale=0.6]{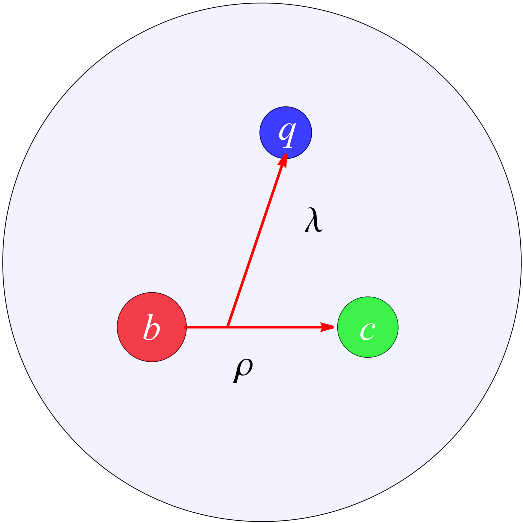}
	\caption{\label{TU1}A typical sketch of the bottom-charmed baryons}
\end{figure}

The bottom-charmed baryon is a three-quark system, where the Jacobi coordinates can be introduced to eliminate the center-of-mass energy.  As illustrated in Figure~\ref{TU1}, the Jacobi coordinates are defined as
\begin{equation}
\boldsymbol{\rho}=\boldsymbol{r_{2}}-\boldsymbol{r_{1}},
\end{equation}
\begin{equation}
\boldsymbol{\lambda}=\boldsymbol{r_{3}}-\frac{{m_{1}}\boldsymbol{r_{1}}+m_{2}\boldsymbol{r_{2}}}{m_{1}+m_{2}},
\end{equation}
\begin{equation}
\boldsymbol{R}=\frac{{m_{1}}\boldsymbol{r_{1}}+m_{2}\boldsymbol{r_{2}}+m_{3}\boldsymbol{r_{3}}}{m_{1}+m_{2}+m_{3}}.
\end{equation}
where $r_{i}$ and $m_{i}$ denote the position vector and the mass of the $i$th quark, respectively. The $\boldsymbol{\rho}$ stands for the relative coordinate between bottom and charm quarks, the $\boldsymbol{\lambda} $ represents the relative coordinate between the light and heavy subsystems, and the $\boldsymbol{R} $ is mass center coordinate.

The bottom-charmed baryons can be divided into four types according to the total wave functions, which are denoted as $\Xi_{bc}$ $\Xi_{bc}^{\prime}$, $\Omega_{bc}$, and $\Omega_{bc}^{\prime}$, respectively. These notations follow those of the singly heavy baryons $\Xi_c$, $\Xi_c^\prime$, $\Xi_b$, and $\Xi_b^\prime$, and the naming scheme can be referred to Refs.~\cite{Chen:2016spr,Eakins:2012jk,Li:2022ywz}.
 In the present work, the orbital wave functions are expanded in terms of a set of Gaussian basis functions that forms an approximate complete set~\cite{Hiyama:2003cu,Hiyama:2018ivm}. Then, the explicit expression for spatial part can be written as 
\begin{equation}\label{total wave}
\Psi(\boldsymbol{\rho},\boldsymbol{\lambda})=\sum_{n,N}C_{n,N}\phi_{n}(\boldsymbol{\rho})\phi_{N}(\boldsymbol{\lambda})
\end{equation}
with
\begin{equation}{\label{wave1}}
\phi_{n}(\boldsymbol{\rho}) =N_{nl_{\rho}}\rho^{l_{\rho}} e^{-\nu_{n}\rho^{2}} Y_{l_{\rho} m_{l_{\rho}}}\left(\boldsymbol{\hat{\rho}}\right)
 \end{equation} 
and
 \begin{equation}{\label{wave2}}
\phi_{N}(\boldsymbol{\lambda}) =N_{Nl_{\lambda}}\lambda^{l_{\lambda}} e^{-\nu_{N}\lambda^{2}} Y_{l_{\lambda} m_{l_{\lambda}}}(\boldsymbol{\hat{\lambda}}),
 \end{equation} 
where $C_{n,N}$ are the expansion coefficients. It is worth noting that the $\rho$ and $\lambda$ mode can be hardly separated in the strict sense for the systems with three non-identical masses. However, in view of the heavy quark symmetry, the orbital wave functions for  bottom-charmed baryons  can be separated approximately and  we prefer to adopt the above trial wave functions to solve the three-body Schr\"odinger equation. 

The range parameters of Gaussian functions are given as 
\begin{equation}
\nu_{n}=1/r_{n}^{2}, r_{n}=r_{1}a^{n-1}(n=1,...,n_{max}),
\end{equation}
\begin{equation}
\nu_{N}=1/R_{N}^{2}, R_{N}=R_{1}A^{N-1}(N=1,...,N_{max}),
\end{equation}
where $n$ and $N$ are the number of Gaussian functions, and $a$ and $A$ are the ratio coefficients. According to the Rayleigh-Ritz variational principle, one can have
\begin{equation}
\sum_{j}^{n_{max} \times N_{max}}[H_{ij}-E N_{ij}]C_{j}=0,(i=1\sim n_{max} \times N_{max}),
 \end{equation}
 where the $H_{ij}$ are the matrix elements in the total color-flavor-spin-orbital bases, $E$ stands for the eigenvalue, and $C_{j}$ are the relevant eigenvectors. Finally, the spectrum of bottom-charmed baryons can be obtained by solving the generalized eigenvalue problem.

 In constructing the total wave functions, we adopt the $j-j$ coupling scheme to investigate the bottom-charmed baryons, where the states are defined as 
\begin{equation}
    \mid J^{P},j\rangle=\mid[(l_{\rho}S_{\rho})_{J_{\rho}}(l_{\lambda}s_{3})_{j}]_{J^{P}}\rangle
\end{equation}
The label $l_{\rho}$ is the orbital quantum number between the two heavy quarks, the $l_{\lambda}$ the orbital quantum number between the light and heavy subsystems, $S_{\rho}$  the spin quantum number of two heavy quarks, $J_{\rho}$  the total angular momentum of heavy quark subsystem, $j$ the total angular momentum of light quark system that is usually known as the light quark spin, and $J^P$ the spin-parity for hadrons. More details of different coupling schemes and their relations can be found in our previous work~\cite{He:2021iwx}. 

 \subsection{Pseudoscalar meson emissions}
\begin{figure}[! htpb]
	\centering
	\includegraphics[scale=0.6]{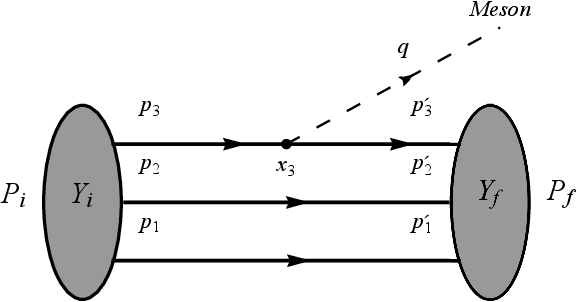}
	\caption{\label{TU2} One meson emission for bottom-charmed baryons
	}
\end{figure}

Besides the mass spectrum, the strong decays can reflect the internal structures of hadrons more explicitly. In this subsection, the approach of strong decays for bottom-charmed baryons is introduced briefly. In the quark model, the pseudoscalar meson can couple to the light quark inside a bottom-charmed baryons through the Yukawa interaction, which is considered to contribute predominantly to one-meson emission decays $Y_{bc}^{i}(P_{i})$ $\to$ $Y_{bc}^{f}(P_{f})$+$M_{p}(q)$ in Figure~\ref{TU2}. The axial-vector coupling between the pseudoscalar meson and a light quark can be written as 
\begin{equation}
 \mathcal{L}_{M_{p}qq}=\frac{g_{A}^{q}}{2 f_{p}} \bar{q} \gamma_{\mu} \gamma_{5} \vec{\tau} q \cdot \partial^{\mu} \vec{M_{p}},   
\end{equation}
where  the $q$ stands for the quark field, $g_{A}^{q}$ is the quark-axial-vector coupling  and $ f_{p}$ is the decay constant. In present work, $f_{\pi}$= 93 $\mathrm{MeV}$and $f_{K}$ =111 $\mathrm{MeV}$ are adopted that have been widely used in quark model calculations~\cite{Arifi:2022ntc,Arifi:2021orx,Nagahiro:2016nsx,Xiao:2017udy,Wang:2018fjm,Liu:2019wdr,Lu:2022puv}.

The wave function for the $Y_{bc}$ $(Y_{bc}=\Xi_{bc}$, $\Xi_{bc}^{\prime} $, $\Omega_{bc}$, and $\Omega_{bc}^{\prime})$ baryon with mass $M_{Y_{bc}}$ in the rest frame can be expressed in the momentum representation  as
\begin{equation}
\begin{aligned}
\left|Y_{bc}(J)\right\rangle=& \sqrt{2 M_{Y_{bc}}} \sum_{\{s, l\}} \int \frac{d^{3} \mathbf{p}_{\rho}}{(2 \pi)^{3}} \int \frac{d^{3} \mathbf{p}_{\lambda}}{(2 \pi)^{3}} \frac{1}{\sqrt{2 m_{1}}} \frac{1}{\sqrt{2 m_{2}}}\\ &\frac{1}{\sqrt{2 m_{3}}} \psi_{l_{\rho}}(\mathbf{p}_{\rho})\psi_{l_{\lambda}}\left(\mathbf{p}_{\lambda}\right)\left|q_{1}\left(p_{1}, s_{1}\right)\right\rangle\left|q_{2}\left(p_{2},s_{2}\right)\right\rangle\\
&\left|q_{3}\left(p_{3}, s_{3}\right)\right\rangle.
\end{aligned}
\end{equation}
Then, the decay amplitude for $Y_{bc}^{i}(P_{i})$ $\to$ $Y_{bc}^{f}(P_{f})$+$M_{p}(q)$ can be obtained by
\begin{equation}{\label{2}}
    \begin{aligned}
-i\mathcal{T}=&-i\frac{g_{A}^{q}g_{f}}{2 f_{p}} \sqrt{2 M_{i}} \sqrt{2 M_{f}}\int d^{3}\lambda e^{i\mathbf{q}_{\lambda}\cdot\mathbf{\lambda}} \\
& \times \Bigg\langle Y_{bc}^{f}\Bigg|i\Bigg\{\Bigg(1-\frac{\omega}{2m_{3}}+\frac{\omega}{m_{1}+m_{2}+m_{3}}\Bigg)\mathbf{\sigma} \cdot \mathbf{q} \\ &+\frac{\omega}{m_{3}}\sigma \cdot \mathbf{p}_{\lambda}\Bigg\}\Bigg| Y_{bc}^{i}\Bigg\rangle,
\end{aligned}
\end{equation}
and  the $\mathbf{q_{\lambda}}$ is defined as 
\begin{equation}
    \mathbf{q_{\lambda}}=\frac{m_{1}+m_{2}}{m_{1}+m_{2}+m_{3}}\mathbf{q}.
\end{equation}
The $g_{f}$ denotes the flavor matrix, $M_{i}$ is the mass of initial baryon, $M_{f}$ is the mass of final baryon,  $q=(\omega,\boldsymbol{q})$ is the 4-momentum of  outgoing pseudoscalar meson, $m_{i}$ is the constituent quark mass with $m_{1}=m_{b}$, $m_{2}=m_{c}$, and $m_{3}=m_{u/d/s}$.

After calculating the mass spectrum, one can  get the wave functions for bottom-charmed baryons, which are applied to estimate the root mean square and the range parameters $\alpha_{\lambda}$. These effective values $\alpha_{\lambda}$ are obtained by equating the
root mean square radius of simple  harmonic oscillator wave functions to that obtained in the nonrelativistic quark model, which has been widely used in the previous studies of strong decays~\cite{Close:2005se,Li:2010vx,Godfrey:2015dia,Godfrey:2015dva,Chen:2016iyi}. Then, the helicity amplitude $\mathcal{A}_{h}$ can be derived from the transition operator and effective parameters $\alpha_{\lambda}$ in the harmonic oscillator wave functions. To calculate the strong decay widths for the pseudoscalar meson emissions, one also need to take into account the phase space factor. Finally, the strong decays for bottom-charmed baryons can be obtained within the helicity bases straightforwardly,
\begin{equation}{\label{decay}}
\Gamma=\frac{1}{4 \pi} \frac{q}{2 M_{i}^{2}} \frac{1}{2 J+1} \sum_{h}\left|\mathcal{A}_{h}\right|^{2}.
\end{equation}

\section{Results and discussion}{\label{low-lying}}
In this section, we first calculated the mass spectrum of bottom-charmed baryons, and then estimate the strong decays for $\lambda-$mode low-lying excited states. Also, we discuss the relativistic corrections for the Roper-like resonances, the mixture of $\lambda-$mode states with the same spin-parity, and the suppression of strong decays for $\rho-$mode and $\rho-\lambda$ hybrid states. In analogy with singly heavy baryons~\cite{Chen:2007xf}, we adopt the symbols $\sim$, $\wedge$,  $\vee$, and $\smile$ on the top of capital
$\Xi$ and $\Omega$ to denote the $\rho-$mode $P-$wave, $\rho-$mode $D-$wave, $\rho-\lambda$ hybrid, and $\rho-$mode radially excited  states, respectively.

\subsection{Mass spectrum}
\begin{figure*}[! htpb]
	\centering
	\includegraphics[scale=0.85]{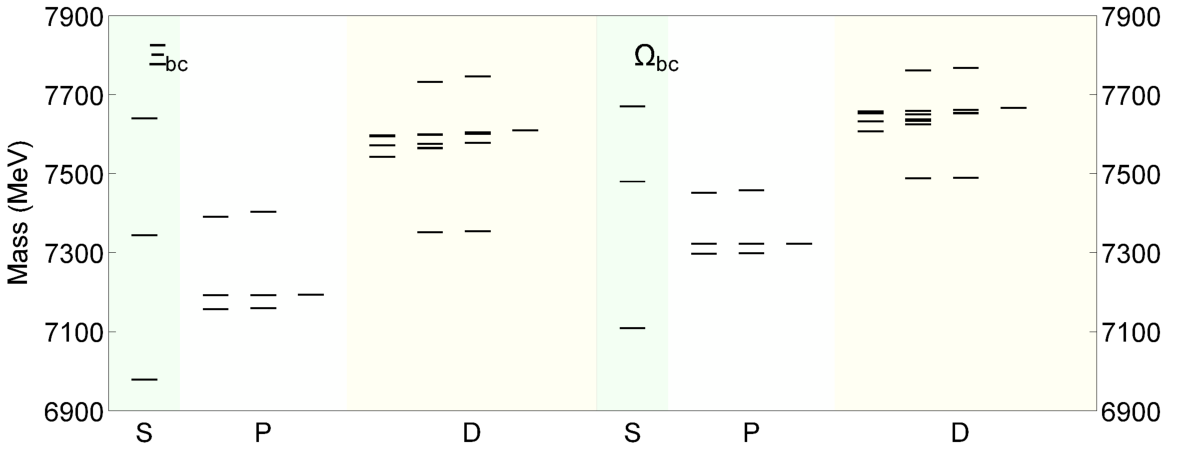}
	\caption{\label{bc.PNG} The calculated mass spectra for  $\Xi_{bc}$ and $\Omega_{bc}$ families}
\end{figure*}

\begin{figure*}[! htpb]
	\centering
	\includegraphics[scale=0.85]{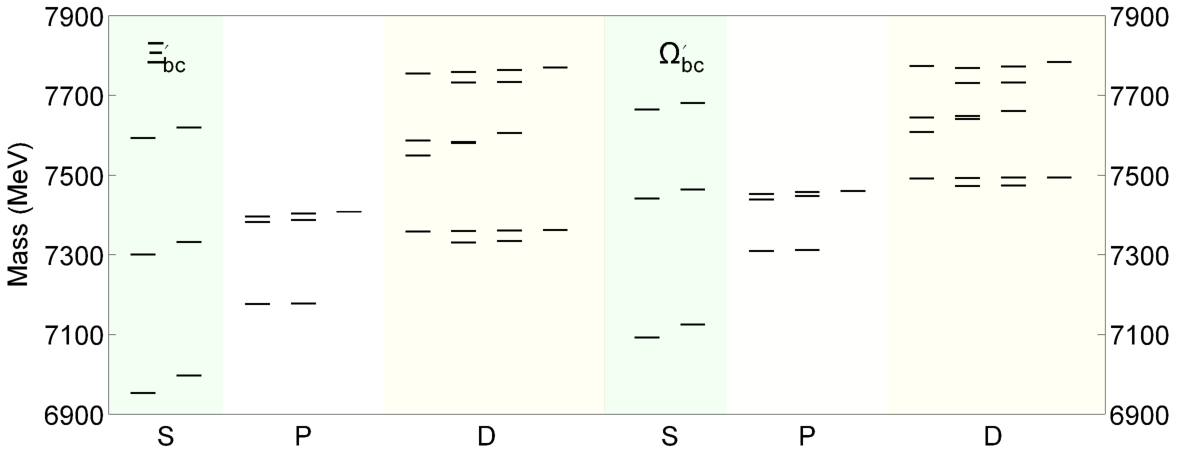}
	\caption{\label{bcp.PNG}The calculated mass spectra for  $\Xi_{bc}^{\prime}$ and $\Omega_{bc}^{\prime}$ families}
\end{figure*}

In the quark model, for the $\Xi_{bc}$ or $\Omega_{bc}$ family, there should be one ground state, five $\rho-$mode $P-$wave  states, two $\lambda-$mode $P-$wave states, two  $\rho-$mode $D-$wave states, two  $\lambda-$mode $D-$wave states,  thirteen $\rho-\lambda$ hybrid $D-$wave states, and two radially excited states; for the $\Xi_{bc}^{\prime}$ or $\Omega_{bc}^{\prime}$ family should exist two ground  states, two $\rho-$mode $P-$wave states, five $\lambda-$mode $P-$wave  states, six  $\rho-$mode $D-$wave states, six   $\lambda-$mode $D-$wave states, five  $\rho-\lambda$ hybrid $D-$wave states, and four radially excited states.  Within the potential model, we obtain the masses and perform them in Table~\ref{mass table 3}, ~\ref{mass table 4}, ~\ref{mass table 1}, and ~\ref{mass table 2}. Also, the full mass spectra are plotted in Figure~\ref{bc.PNG} and \ref{bcp.PNG} for reference.

\begin{table}[!htbp]
	\begin{center}
		\caption{\label{mass table 3} The mass spectrum of $\Xi_{bc}$ in MeV.}
		\renewcommand{\arraystretch}{1.2}
		\normalsize
		\begin{tabular*}{8cm}{@{\extracolsep{\fill}}p{1.4cm}<{\centering}p{0.2cm}<{\centering}p{0.2cm}<{\centering}p{0.2cm}<{\centering}p{0.2cm}<{\centering}p{0.2cm}<{\centering}p{0.2cm}<{\centering}p{0.2cm}<{\centering}p{0.2cm}<{\centering}p{0.8cm}<{\centering}p{0.8cm}<{\centering}}
			\hline\hline
			States& $n_{\rho}$ & $n_{\lambda}$ & $l_{\rho}$ & $l_{\lambda}$ & $S_{\rho}$ & $J_{\rho}$ & $j$	&  $J^P$&$\alpha_{\lambda}$&Mass\\
			\hline
			$\Xi_{bc}(1S)$	  &	0	&	0	&	0	&	0	&	1   &1    &$\frac{1}{2}$	&	$\frac{1}{2}^+$ & 347 & 6979 \\ 
			$\breve{\Xi}_{bc} (2S)$	  &	1	&	0	&	0	&	0	&	1   &1    &$\frac{1}{2}$	&	$\frac{1}{2}^+$ &507  & 7344 \\
			$\Xi_{bc} (2S)$	  &	0	&	1	&	0	&	0	&	1   &1    &$\frac{1}{2}$	&	$\frac{1}{2}^+$ & 489 & 7640 \\
			$\tilde{\Xi}_{bc} (\frac{1}{2}^-, \frac{1}{2})$&	0	&	0	&	1	&	0	&	1   &   0  &  $\frac{1}{2}$	&   $\frac{1}{2}^-$ &438 &	7157 \\
			$\tilde{\Xi}_{bc} (\frac{3}{2}^-, \frac{1}{2})$&	0	&	0	&	1	&	0	&	1   &   1  &  $\frac{1}{2}$	&   $\frac{3}{2}^-$ &437 &	7159 \\
			$\tilde{\Xi}_{bc} (\frac{1}{2}^-, \frac{1}{2})$&	0	&	0	&	1	&	0	&	1   &   1  &  $\frac{1}{2}$	&   $\frac{1}{2}^-$ &441 &	7192 \\
			$\tilde{\Xi}_{bc} (\frac{3}{2}^-, \frac{1}{2})$&	0	&	0	&	1	&	0	&	1   &   2  &  $\frac{1}{2}$	&   $\frac{3}{2}^-$ & 440&	7192 \\
			$\tilde{\Xi}_{bc} (\frac{5}{2}^-, \frac{1}{2})$&    0	&	0	&	1	&	0	&	1   &   2  &  $\frac{1}{2}$	&	$\frac{5}{2}^-$ &432 &	7193 \\
			$\Xi_{bc} (\frac{1}{2}^-, \frac{1}{2})$&	0	&	0	&	0	&	1	&	0   &	0  &  $\frac{1}{2}$	&   $\frac{1}{2}^-$ &305 &	7391\\
			$\Xi_{bc} (\frac{3}{2}^-, \frac{3}{2})$&	0	&	0	&	0	&	1	&	0   &	0  &  $\frac{3}{2}$	&	$\frac{3}{2}^-$ &302 &	7403 \\
			$\hat{\Xi}_{bc} (\frac{3}{2}^+, \frac{1}{2})$&	0	&	0	&	2	&	0	&	0   &   2  &  $\frac{1}{2}$	&   $\frac{3}{2}^+$ &501 &	7352 \\
			$\hat{\Xi}_{bc} (\frac{5}{2}^+, \frac{1}{2})$&	0	&	0	&	2	&	0	&	0   &   2  &  $\frac{1}{2}$	&   $\frac{5}{2}^+$ & 500&	7354 \\
			$\Xi_{bc} (\frac{3}{2}^+, \frac{3}{2})$&	0	&	0	&	0	&	2	&	0   &   0  &  $\frac{3}{2}$	&   $\frac{3}{2}^+$ &291 &	7732 \\
			$\Xi_{bc} (\frac{5}{2}^+, \frac{5}{2})$&	0	&	0	&	0	&	2	&	0   &   0  &  $\frac{5}{2}$	&   $\frac{5}{2}^+$ &282 &	7746 \\
			$\check{\Xi}_{bc} (\frac{1}{2}^+, \frac{1}{2})$&	0	&	0	&	1	&	1	&	1   &   0  &  $\frac{1}{2}$	&   $\frac{1}{2}^+$ &368&	7543 \\
			$\check{\Xi}_{bc} (\frac{3}{2}^+, \frac{3}{2})$&	0	&	0	&	1	&	1	&	1   &   0  &  $\frac{3}{2}$	&	$\frac{3}{2}^+$ &367&	7564\\
			$\check{\Xi}_{bc} (\frac{1}{2}^+, \frac{1}{2})$&	0	&	0	&	1	&	1	&	1   &   1  &  $\frac{1}{2}$	&   $\frac{1}{2}^+$ &366 &	7572 \\
			$\check{\Xi}_{bc} (\frac{3}{2}^+, \frac{1}{2})$&	0	&	0	&	1	&	1	&	1   &   1  &  $\frac{1}{2}$	&   $\frac{3}{2}^+$ &366 &	7575 \\
			$\check{\Xi}_{bc} (\frac{1}{2}^+, \frac{3}{2})$&	0	&	0	&	1	&	1	&	1   &   1  &  $\frac{3}{2}$	&   $\frac{1}{2}^+$ &359 &	7597 \\
			$\check{\Xi}_{bc} (\frac{3}{2}^+, \frac{3}{2})$&	0	&	0	&	1	&	1	&	1   &   1  &  $\frac{3}{2}$	&   $\frac{3}{2}^+$ &359 &	7599 \\
			$\check{\Xi}_{bc} (\frac{5}{2}^+, \frac{3}{2})$&	0	&	0	&	1	&	1	&	1   &   1  &  $\frac{3}{2}$	&	$\frac{5}{2}^+$ &352 &	7604\\
			$\check{\Xi}_{bc} (\frac{3}{2}^+, \frac{1}{2})$&	0	&	0	&	1	&	1	&	1   &   2  &  $\frac{1}{2}$	&	$\frac{3}{2}^+$ &367 &	7565\\
			$\check{\Xi}_{bc} (\frac{5}{2}^+, \frac{1}{2})$&	0	&	0	&	1	&	1	&	0   &   2  &  $\frac{1}{2}$	&	$\frac{5}{2}^+$ &366 &	7578\\
			$\check{\Xi}_{bc} (\frac{1}{2}^+, \frac{3}{2})$&	0	&	0	&	1	&	1	&	0   &   2  &  $\frac{3}{2}$	&	$\frac{1}{2}^+$ &353 &	7595\\
			$\check{\Xi}_{bc} (\frac{3}{2}^+, \frac{3}{2})$&	0	&	0	&	1	&	1	&	0   &   2  &  $\frac{3}{2}$	&	$\frac{3}{2}^+$ &352 &	7598\\
			$\check{\Xi}_{bc} (\frac{5}{2}^+, \frac{3}{2})$&	0	&	0	&	1	&	1	&	0   &   2  &  $\frac{3}{2}$	&	$\frac{5}{2}^+$ &352&	7601\\
			$\check{\Xi}_{bc} (\frac{7}{2}^+, \frac{3}{2})$&	0	&	0	&	1	&	1	&	0   &   2  &  $\frac{3}{2}$	&	$\frac{7}{2}^+$ & 350&	7609\\
			\hline\hline
		\end{tabular*}
	\end{center}
\end{table}

\begin{table}[!htbp]
	\begin{center}
		\caption{\label{mass table 4} The mass spectrum of $\Omega_{bc}$ in MeV.}
		\renewcommand{\arraystretch}{1.2}
		\normalsize
		\begin{tabular*}{8cm}{@{\extracolsep{\fill}}p{1.4cm}<{\centering}p{0.2cm}<{\centering}p{0.2cm}<{\centering}p{0.2cm}<{\centering}p{0.2cm}<{\centering}p{0.2cm}<{\centering}p{0.2cm}<{\centering}p{0.2cm}<{\centering}p{0.2cm}<{\centering}p{0.8cm}<{\centering}p{0.8cm}<{\centering}}
			\hline\hline
			States& $n_{\rho}$ & $n_{\lambda}$ & $l_{\rho}$ & $l_{\lambda}$ & $S_{\rho}$ & $J_{\rho}$ & $j$	&  $J^P$&$\alpha_{\lambda}$&Mass\\
			\hline
			$\Omega_{bc}(1S)$	  &	0	&	0	&	0	&	0	&	1   &1    &$\frac{1}{2}$	&	$\frac{1}{2}^+$ & 419 & 7109 \\ 
			$\breve{\Omega}_{bc} (2S)$	  &	1	&	0	&	0	&	0	&	1   &1    &$\frac{1}{2}$	&	$\frac{1}{2}^+$  &601 & 7480 \\
			$\Omega_{bc} (2S)$	  &	0	&	1	&	0	&	0	&	1   &1    &$\frac{1}{2}$	&	$\frac{1}{2}^+$  &351 & 7670 \\
			$\tilde{\Omega}_{bc} (\frac{1}{2}^-, \frac{1}{2})$&	0	&	0	&	1	&	0	&	1   &   0  &  $\frac{1}{2}$	&   $\frac{1}{2}^-$  &506&	7297 \\
			$\tilde{\Omega}_{bc} (\frac{3}{2}^-, \frac{1}{2})$&	0	&	0	&	1	&	0	&	1   &   1  &  $\frac{1}{2}$	&   $\frac{3}{2}^-$  &506&7298 \\
			$\tilde{\Omega}_{bc} (\frac{1}{2}^-, \frac{1}{2})$&	0	&	0	&	1	&	0	&	1   &   1  &  $\frac{1}{2}$	&   $\frac{1}{2}^-$  &500	&7322\\
			$\tilde{\Omega}_{bc} (\frac{3}{2}^-, \frac{1}{2})$&	0	&	0	&	1	&	0	&	1   &   2  &  $\frac{1}{2}$	&   $\frac{3}{2}^-$  &500&7322 \\
			$\tilde{\Omega}_{bc} (\frac{5}{2}^-, \frac{1}{2})$& 0	&	0	&	1	&	0	&	1   &   2  &  $\frac{1}{2}$	&	$\frac{5}{2}^-$  &500	&7323 \\
			$\Omega_{bc} (\frac{1}{2}^-, \frac{1}{2})$&	0	&	0	&	0	&	1	&	0   &	0  &  $\frac{1}{2}$	&   $\frac{1}{2}^-$  &362	&7451 \\
			$\Omega_{bc} (\frac{3}{2}^-, \frac{3}{2})$&	0	&	0	&	0	&	1	&	0   &	0  &  $\frac{3}{2}$	&	$\frac{3}{2}^-$  &360	&7458 \\
			$\hat{\Omega}_{bc} (\frac{3}{2}^+, \frac{1}{2})$&	0	&	0	&	2	&	0	&	0   &   2  &  $\frac{1}{2}$	&  $\frac{3}{2}^+$  &583	&7488 \\
			$\hat{\Omega}_{bc} (\frac{5}{2}^+, \frac{1}{2})$&	0	&	0	&	2	&	0	&	0   &   2  &  $\frac{1}{2}$	&   $\frac{5}{2}^+$  &582	&7489 \\
			$\Omega_{bc} (\frac{3}{2}^+, \frac{3}{2})$&	0	&	0	&	0	&	2	&	0   &   0  &  $\frac{3}{2}$	&   $\frac{3}{2}^+$  &306	&7761 \\
			$\Omega_{bc} (\frac{5}{2}^+, \frac{5}{2})$&	0	&	0	&	0	&	2	&	0   &   0  &  $\frac{5}{2}$	&   $\frac{5}{2}^+$  &305&7768 \\
			$\check{\Omega}_{bc} (\frac{1}{2}^+, \frac{1}{2})$&	0	&	0	&	1	&	1	&	1   &   0  &  $\frac{1}{2}$	&   $\frac{1}{2}+$  &431&7607 \\
			$\check{\Omega}_{bc} (\frac{3}{2}^+, \frac{3}{2})$&	0	&	0	&	1	&	1	&	1   &   0  &  $\frac{3}{2}$	&	$\frac{3}{2}^+$  &422&7625 \\
			$\check{\Omega}_{bc} (\frac{1}{2}^+, \frac{1}{2})$&	0	&	0	&	1	&	1	&	1   &   1  &  $\frac{1}{2}$	&   $\frac{1}{2}^+$  &417	&7632 \\
			$\check{\Omega}_{bc} (\frac{3}{2}^+, \frac{1}{2})$&	0	&	0	&	1	&	1	&	1   &   1  &  $\frac{1}{2}$	&   $\frac{3}{2}^+$  &417	&7634 \\
			$\check{\Omega}_{bc} (\frac{1}{2}^+, \frac{3}{2})$&	0	&	0	&	1	&	1	&	1   &   1  &  $\frac{3}{2}$	&   $\frac{1}{2}^+$  &	413&7652 \\
			$\check{\Omega}_{bc} (\frac{3}{2}^+, \frac{3}{2})$&	0	&	0	&	1	&	1	&	1   &   1  &  $\frac{3}{2}$	&   $\frac{3}{2}^+$  &	413&7650 \\
			$\check{\Omega}_{bc} (\frac{5}{2}^+, \frac{3}{2})$&	0	&	0	&	1	&	1	&	1   &   1  &  $\frac{3}{2}$	&	$\frac{5}{2}^+$  &412	&7654 \\
			$\check{\Omega}_{bc} (\frac{3}{2}^+, \frac{1}{2})$&	0	&	0	&	1	&	1	&	1   &   2  &  $\frac{1}{2}$	&	$\frac{3}{2}^+$  &  417&7637 \\
			$\check{\Omega}_{bc} (\frac{5}{2}^+, \frac{1}{2})$&	0	&	0	&	1	&	1	&	0   &   2  &  $\frac{1}{2}$	&	$\frac{5}{2}^+$  &412	&7652 \\
			$\check{\Omega}_{bc} (\frac{1}{2}^+, \frac{3}{2})$&	0	&	0	&	1	&	1	&	0   &   2  &  $\frac{3}{2}$	&	$\frac{1}{2}^+$  &	411&7658 \\
			$\check{\Omega}_{bc} (\frac{3}{2}^+, \frac{3}{2})$&	0	&	0	&	1	&	1	&	0   &   2  &  $\frac{3}{2}$	&	$\frac{3}{2}^+$  &411	&7659 \\
			$\check{\Omega}_{bc} (\frac{5}{2}^+, \frac{3}{2})$&	0	&	0	&	1	&	1	&	0   &   2  &  $\frac{3}{2}$	&	$\frac{5}{2}^+$  &	409&7661 \\
			$\check{\Omega}_{bc} (\frac{7}{2}^+, \frac{3}{2})$&	0	&	0	&	1	&	1	&	0   &   2  &  $\frac{3}{2}$	&	$\frac{7}{2}^+$  &405	&7667 \\
			\hline\hline
		\end{tabular*}
	\end{center}
\end{table}

\begin{table}[!htbp]
	\begin{center}
		\caption{\label{mass table 1} The mass spectrum of $\Xi_{bc}^{\prime}$ in MeV.}
		\renewcommand{\arraystretch}{1.2}
		\normalsize
		\begin{tabular*}{8cm}{@{\extracolsep{\fill}}p{1.4cm}<{\centering}p{0.2cm}<{\centering}p{0.2cm}<{\centering}p{0.2cm}<{\centering}p{0.2cm}<{\centering}p{0.2cm}<{\centering}p{0.2cm}<{\centering}p{0.2cm}<{\centering}p{0.2cm}<{\centering}p{0.8cm}<{\centering}p{0.8cm}<{\centering}}
			\hline\hline
			States& $n_{\rho}$ & $n_{\lambda}$ & $l_{\rho}$ & $l_{\lambda}$ & $S_{\rho}$ & $J_{\rho}$ & $j$	&  $J^P$&$\alpha_{\lambda}$&Mass\\
			\hline
			$\Xi_{bc}^{\prime} (1S)$	  &	0	&	0	&	0	&	0	&	1   &1    &$\frac{1}{2}$	&	$\frac{1}{2}^+$ & 373 & 6953 \\
			$\Xi_{bc}^{\prime*}(1S)$      &	0	&	0	&	0	&	0	&	1   &1    &$\frac{1}{2}$	&	$\frac{3}{2}^+$ & 357   & 6997 \\ 
			$\breve{\Xi}_{bc}^{\prime} (2S)$	  &	1	&	0	&	0	&	0	&	1   &1    &$\frac{1}{2}$	&	$\frac{1}{2}^+$ & 533  & 7301 \\
			$\breve{\Xi}_{bc}^{\prime*}(2S)$      &	1	&	0	&	0	&	0	&	1   &1    &$\frac{1}{2}$	&	$\frac{3}{2}^+$ & 520 &	7332 \\
			$\Xi_{bc}^{\prime} (2S)$	  &	0	&	1	&	0	&	0	&	1   &1    &$\frac{1}{2}$	&	$\frac{1}{2}^+$ & 508  & 7593 \\
			$\Xi_{bc}^{\prime*}(2S)$      &	0	&	1	&	0	&	0	&	1   &1    &$\frac{3}{2}$	&	$\frac{3}{2}^+$ & 501  &	7619 \\
			$\tilde{\Xi}_{bc}^{\prime} (\frac{1}{2}^-, \frac{1}{2})$&	0	&	0	&	1	&	0	&	0   &   1  &  $\frac{1}{2}$	&   $\frac{1}{2}^-$ & 438 &	7176 \\
			$\tilde{\Xi}_{bc}^{\prime} (\frac{3}{2}^-, \frac{1}{2})$&	0	&	0	&	1	&	0	&	0   &   1  &  $\frac{1}{2}$	&   $\frac{3}{2}^-$ &437 &	7178 \\
			$\Xi_{bc}^{\prime} (\frac{1}{2}^-, \frac{1}{2})$&	0	&	0	&	0	&	1	&	1   &   1  &  $\frac{1}{2}$	&   $\frac{1}{2}^-$ & 319 &	7382 \\
			$\Xi_{bc}^{\prime} (\frac{3}{2}^-, \frac{1}{2})$&	0	&	0	&	0	&	1	&	1   &   1  &  $\frac{1}{2}$	&   $\frac{3}{2}^-$ & 315 &	7387 \\
			$\Xi_{bc}^{\prime} (\frac{1}{2}^-, \frac{3}{2})$&    0	&	0	&	0	&	1	&	1   &   1  &  $\frac{3}{2}$	&	$\frac{1}{2}^-$ & 313 &	7396 \\
			$\Xi_{bc}^{\prime} (\frac{3}{2}^-, \frac{3}{2})$&	0	&	0	&	0	&	1	&	1   &	1  &  $\frac{3}{2}$	&   $\frac{3}{2}^-$ & 311 &	7404 \\
			$\Xi_{bc}^{\prime} (\frac{5}{2}^-, \frac{3}{2})$&	0	&	0	&	0	&	1	&	1   &	1  &  $\frac{3}{2}$	&	$\frac{5}{2}^-$ &310 &	7408 \\
			$\hat{\Xi}_{bc}^{\prime} (\frac{3}{2}^+, \frac{1}{2})$&	0	&	0	&	2	&	0	&	1   &   2  &  $\frac{1}{2}$	&   $\frac{3}{2}^+$ &514 &	7331 \\
			$\hat{\Xi}_{bc}^{\prime} (\frac{5}{2}^+, \frac{1}{2})$&	0	&	0	&	2	&	0	&	1   &   2  &  $\frac{1}{2}$	&   $\frac{5}{2}^+$ &514 &	7334 \\
			$\hat{\Xi}_{bc}^{\prime} (\frac{1}{2}^+, \frac{1}{2})$&	0	&	0	&	2	&	0	&	1   &   1  &  $\frac{1}{2}$	&   $\frac{1}{2}^+$ &503 &	7359 \\
			$\hat{\Xi}_{bc}^{\prime} (\frac{3}{2}^+, \frac{1}{2})$&	0	&	0	&	2	&	0	&	1   &   1  &  $\frac{1}{2}$	&   $\frac{3}{2}^+$ &502 &	7360 \\
			$\hat{\Xi}_{bc}^{\prime} (\frac{5}{2}^+, \frac{1}{2})$&	0	&	0	&	2	&	0	&	1   &   3  &  $\frac{1}{2}$	&   $\frac{5}{2}^+$ &502 &	7361 \\
			$\hat{\Xi}_{bc}^{\prime} (\frac{7}{2}^+, \frac{1}{2})$&	0	&	0	&	2	&	0	&	1   &   3  &  $\frac{1}{2}$	&	$\frac{7}{2}^+$ &502&	7362 \\
			$\Xi_{bc}^{\prime} (\frac{3}{2}^+, \frac{3}{2})$&	0	&	0	&	0	&	2	&	1   &   1  &  $\frac{3}{2}$	&   $\frac{3}{2}^+$ & 295 &	7732 \\
			$\Xi_{bc}^{\prime} (\frac{5}{2}^+, \frac{3}{2})$&	0	&	0	&	0	&	2	&	1   &   1  &  $\frac{3}{2}$	&   $\frac{5}{2}^+$ & 295 &	7733\\
			$\Xi_{bc}^{\prime} (\frac{1}{2}^+, \frac{3}{2})$&	0	&	0	&	0	&	2	&	1   &   1  &  $\frac{3}{2}$	&   $\frac{1}{2}^+$ & 280 &	7755 \\
			$\Xi_{bc}^{\prime} (\frac{3}{2}^+, \frac{5}{2})$&	0	&	0	&	0	&	2	&	1   &   1  &  $\frac{5}{2}$	&   $\frac{3}{2}^+$ & 279 &	7759 \\
			$\Xi_{bc}^{\prime} (\frac{5}{2}^+, \frac{5}{2})$&	0	&	0	&	0	&	2	&	1   &   1  &  $\frac{5}{2}$	&	$\frac{5}{2}^+$ & 278 &	7764\\
			$\Xi_{bc}^{\prime} (\frac{7}{2}^+, \frac{5}{2})$&	0	&	0	&	0	&	2	&	1   &   1  &  $\frac{5}{2}$	&	$\frac{7}{2}^+$ & 276 &	7770\\
			$\check{\Xi}_{bc}^{\prime} (\frac{1}{2}^+, \frac{1}{2})$&	0	&	0	&	1	&	1	&	0   &   1  &  $\frac{1}{2}$	&	$\frac{1}{2}^+$ & 388&	7549\\
			$\check{\Xi}_{bc}^{\prime} (\frac{3}{2}^+, \frac{1}{2})$&	0	&	0	&	1	&	1	&	0   &   1  &  $\frac{1}{2}$	&	$\frac{3}{2}^+$ &370 &	7581\\
			$\check{\Xi}_{bc}^{\prime} (\frac{1}{2}^+, \frac{3}{2})$&	0	&	0	&	1	&	1	&	0   &   1  &  $\frac{3}{2}$	&	$\frac{1}{2}^+$ & 368&	7587\\
			$\check{\Xi}_{bc}^{\prime} (\frac{3}{2}^+, \frac{3}{2})$&	0	&	0	&	1	&	1	&	0   &   1  &  $\frac{3}{2}$	&	$\frac{3}{2}^+$ &369 &	7583\\
			$\check{\Xi}_{bc}^{\prime} (\frac{5}{2}^+, \frac{3}{2})$&	0	&	0	&	1	&	1	&	0   &   1  &  $\frac{3}{2}$	&	$\frac{5}{2}^+$ & 350&	7606\\
			\hline\hline
		\end{tabular*}
	\end{center}
\end{table}
\begin{table}[!htbp]
	\begin{center}
		\caption{\label{mass table 2} The mass spectrum of $\Omega_{bc}^{\prime}$ in MeV.}
		\renewcommand{\arraystretch}{1.2}
		\normalsize
		\begin{tabular*}{8cm}{@{\extracolsep{\fill}}p{1.4cm}<{\centering}p{0.2cm}<{\centering}p{0.2cm}<{\centering}p{0.2cm}<{\centering}p{0.2cm}<{\centering}p{0.2cm}<{\centering}p{0.2cm}<{\centering}p{0.2cm}<{\centering}p{0.2cm}<{\centering}p{0.8cm}<{\centering}p{0.8cm}<{\centering}}
			\hline\hline
			States& $n_{\rho}$ & $n_{\lambda}$ & $l_{\rho}$ & $l_{\lambda}$ & $S_{\rho}$ & $J_{\rho}$ & $j$	&  $J^P$&$\alpha_{\lambda}$&Mass\\
			\hline
			$\Omega_{bc}^{\prime} (1S)$	  &	0	&	0	&	0	&	0	&	1   &1    &$\frac{1}{2}$	&	$\frac{1}{2}^+$ &455  & 7092 \\
			$\Omega_{bc}^{\prime*}(1S)$      &	0	&	0	&	0	&	0	&	1   &1    &$\frac{1}{2}$	&	$\frac{3}{2}^+$ &397  & 7125 \\ 
			$\breve{\Omega}_{bc}^{\prime} (2S)$	  &	1	&	0	&	0	&	0	&	1   &1    &$\frac{1}{2}$	&	$\frac{1}{2}^+$  &635 & 7441 \\
			$\breve{\Omega}_{bc}^{\prime*}(2S)$      &	1	&	0	&	0	&	0	&	1   &1    &$\frac{1}{2}$	&	$\frac{3}{2}^+$ & 622 &	7464 \\
			$\Omega_{bc}^{\prime} (2S)$	  &	0	&	1	&	0	&	0	&	1   &1    &$\frac{1}{2}$	&	$\frac{1}{2}^+$  &345 & 7664 \\
			$\Omega_{bc}^{\prime*}(2S)$      &	0	&	1	&	0	&	0	&	1   &1    &	$\frac{1}{2}$	&	$\frac{3}{2}^+$ &343  &	7681 \\
			$\tilde{\Omega}_{bc}^{\prime} (\frac{1}{2}^-, \frac{1}{2})$&	0	&	0	&	1	&	0	&	0   &   1  &  $\frac{1}{2}$	&    $\frac{1}{2}^-$ &505 &	7310 \\
			$\tilde{\Omega}_{bc}^{\prime} (\frac{3}{2}^-, \frac{1}{2})$&	0	&	0	&	1	&	0	&	0   &   1  &  $\frac{1}{2}$	&   $\frac{3}{2}^-$ &504 &	7312 \\
			$\Omega_{bc}^{\prime} (\frac{1}{2}^-, \frac{1}{2})$&	0	&	0	&	0	&	1	&	1   &   1  &  $\frac{1}{2}$	&   $\frac{1}{2}^-$  &374 &	7439 \\
			$\Omega_{bc}^{\prime} (\frac{3}{2}^-, \frac{1}{2})$&	0	&	0	&	0	&	1	&	1   &   1  &  $\frac{1}{2}$	&   $\frac{3}{2}^-$ &369 &	7448 \\
			$\Omega_{bc}^{\prime} (\frac{1}{2}^-, \frac{3}{2})$&    0	&	0	&	0	&	1	&	1   &   1  &  $\frac{3}{2}$	&	$\frac{1}{2}^-$ &366 &	7453 \\
			$\Omega_{bc}^{\prime} (\frac{3}{2}^-, \frac{3}{2})$&	0	&	0	&	0	&	1	&	1   &	1  &  $\frac{3}{2}$	&   $\frac{3}{2}^-$ &365 &	7458 \\
			$\Omega_{bc}^{\prime} (\frac{5}{2}^-, \frac{3}{2})$&	0	&	0	&	0	&	1	&	1   &	1  &  $\frac{3}{2}$	&	$\frac{5}{2}^-$ &364 &	7460 \\
			$\hat{\Omega}_{bc}^{\prime} (\frac{3}{2}^+, \frac{1}{2})$&	0	&	0	&	2	&	0	&	1   &   2  &  $\frac{1}{2}$	&   $\frac{3}{2}^+$ &590 &	7472 \\
			$\hat{\Omega}_{bc}^{\prime} (\frac{5}{2}^+, \frac{1}{2})$&	0	&	0	&	2	&	0	&	1   &   2  &  $\frac{1}{2}$	&   $\frac{5}{2}^+$ &589 &	7474 \\
			$\hat{\Omega}_{bc}^{\prime} (\frac{1}{2}^+, \frac{1}{2})$&	0	&	0	&	2	&	0	&	1   &   1  &  $\frac{1}{2}$	&   $\frac{1}{2}^+$ &583 &	7492 \\
			$\hat{\Omega}_{bc}^{\prime} (\frac{3}{2}^+, \frac{1}{2})$&	0	&	0	&	2	&	0	&	1   &   1  &  $\frac{1}{2}$	&   $\frac{3}{2}^+$ &583 &	7493 \\
			$\hat{\Omega}_{bc}^{\prime} (\frac{5}{2}^+, \frac{1}{2})$&	0	&	0	&	2	&	0	&	1   &   3  &  $\frac{1}{2}$	&   $\frac{5}{2}^+$ & 582&	7494 \\
			$\hat{\Omega}_{bc}^{\prime} (\frac{7}{2}^+, \frac{1}{2})$&	0	&	0	&	2	&	0	&	1   &   3  &  $\frac{1}{2}$	&	$\frac{7}{2}^+$ &582 &	7494 \\
			$\hat{\Omega}_{bc}^{\prime} (\frac{3}{2}^+, \frac{3}{2})$&	0	&	0	&	0	&	2	&	1   &   1  &  $\frac{3}{2}$	&   $\frac{3}{2}^+$ &365 &	7731 \\
			$\Omega_{bc}^{\prime} (\frac{5}{2}^+, \frac{3}{2})$&	0	&	0	&	0	&	2	&	1   &   1  &  $\frac{3}{2}$	&   $\frac{5}{2}^+$ &365 &	7732 \\
			$\Omega_{bc}^{\prime} (\frac{1}{2}^+, \frac{3}{2})$&	0	&	0	&	0	&	2	&	1   &   1  &  $\frac{3}{2}$	&   $\frac{1}{2}^+$ &348 &	7774 \\
			$\Omega_{bc}^{\prime} (\frac{3}{2}^+, \frac{5}{2})$&	0	&	0	&	0	&	2	&	1   &   1  &  $\frac{5}{2}$	&   $\frac{3}{2}^+$ &349 &	7769 \\
			$\Omega_{bc}^{\prime} (\frac{5}{2}^+, \frac{5}{2})$&	0	&	0	&	0	&	2	&	1   &   1  &  $\frac{5}{2}$	&	$\frac{5}{2}^+$ &348 &	7773 \\
			$\Omega_{bc}^{\prime} (\frac{7}{2}^+, \frac{5}{2})$&	0	&	0	&	0	&	2	&	1   &   1  &  $\frac{5}{2}$	&	$\frac{7}{2}^+$ &343 &	7784 \\
			$\check{\Omega}_{bc}^{\prime} (\frac{1}{2}^+, \frac{1}{2})$&	0	&	0	&	1	&	1	&	0   &   1  &  $\frac{1}{2}$	&	$\frac{1}{2}^+$ &445 &	7608 \\
			$\check{\Omega}_{bc}^{\prime} (\frac{3}{2}^+, \frac{1}{2})$&	0	&	0	&	1	&	1	&	0   &   1  &  $\frac{1}{2}$	&	$\frac{3}{2}^+$ &438 &	7641 \\
			$\check{\Omega}_{bc}^{\prime} (\frac{1}{2}^+, \frac{3}{2})$&	0	&	0	&	1	&	1	&	0   &   1  &  $\frac{3}{2}$	&	$\frac{1}{2}^+$ &437 &	7645 \\
			$\check{\Omega}_{bc}^{\prime} (\frac{3}{2}^+, \frac{3}{2})$&	0	&	0	&	1	&	1	&	0   &   1  &  $\frac{3}{2}$	&	$\frac{3}{2}^+$ &433 &	7648 \\
			$\check{\Omega}_{bc}^{\prime} (\frac{5}{2}^+, \frac{3}{2})$&	0	&	0	&	1	&	1	&	0   &   1  &  $\frac{3}{2}$	&	$\frac{5}{2}^+$ &424 &	7661 \\
			\hline\hline
		\end{tabular*}
	\end{center}
\end{table}

\begin{figure*}[! htpb]
	\centering
	\includegraphics[scale=0.75]{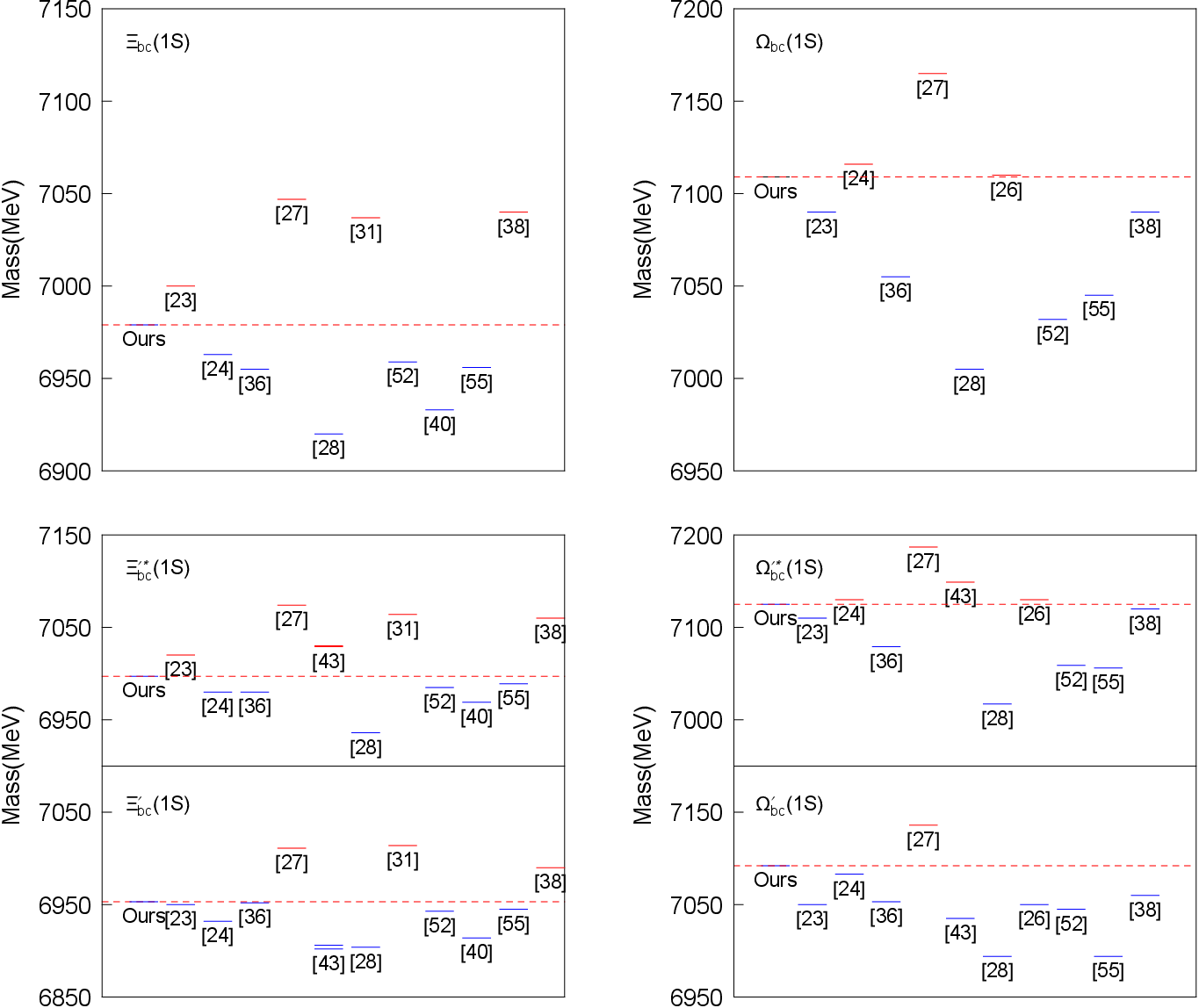}
	\caption{\label{Groundvs}A comparison of the ground states for bottom-charmed baryons from various model predictions.}
\end{figure*}

We first refer to the ground states for bottom-charmed baryons. From the Table~\ref{mass table VS2} and Figure~\ref{Groundvs}, The lowest $\Xi_{bc}$, $\Xi_{bc}^\prime$, $\Omega_{bc}$, and $\Omega_{bc}^\prime$ states are predicted to be about 6979, 6953, 7109, and 7092 MeV, respectively. It can be seen that our results are consistent with some works~\cite{Ebert:1996ec,Ebert:2002ig,Li:2022ywz}, while differ with others about $50 \sim 100$ MeV~\cite{Roberts:2007ni,Oudichhya:2022ssc,Giannuzzi:2009gh}. In the singly heavy baryons $\Xi_{c/b}^{(\prime)}$, the lowest states are $\Xi_{c/b}$ rather than  $\Xi_{c/b}^{\prime}$. However, for the bottom-charmed baryons, the lowest states are $\Xi_{bc}^{\prime}$ and $\Omega_{bc}^{\prime}$, which is totally different with singly heavy sector. This is due to the different masses in spin-spin term $V^{SS}\left(r_{ij}\right)$ in the nonrelativistic quark model for singly and doubly heavy baryons, where the pairwise interactions among three quarks together with spin wave functions compete with each other. Moreover, the mass splittings for $\Xi_{bc}^{\prime*}-\Xi_{bc}^{\prime}$ and $\Omega_{bc}^{\prime*}-\Omega_{bc}^{\prime}$ are 44 and 33 MeV, respectively. These quite small mass gaps are  caused by spin-spin interaction which is inversely proportional to quark masses.  Because of this, 
the pion emission between the ground states for bottom-charmed baryons is prohibited and only decay mode is by electroweak processes, which is in contrast with the strange sector allowing the strong decay $\Xi^{*}\to\Xi\pi$. In particular, For the lowest $\Xi_{bc}^{(\prime)}$ and $\Omega_{bc}^{(\prime)}$ states, future experiments can search for them in via $b \to c$ weak transitions.

 \begin{table*}[htbp]
	\caption{\label{mass table VS2} Masses of ground states for $\Xi_{bc}^{(\prime)}$and $\Omega_{bc}^{(\prime)}$ baryons compared with different calculations. The units are in MeV.}
	\begin{ruledtabular}
		\begin{tabular}{ccccccccccccccc}
			&States
			&$J^{P}$
			&Our work
			&\cite{Ebert:1996ec}
			&\cite{Ebert:2002ig}
			&\cite{Li:2022ywz}
			&\cite{Roberts:2007ni}
			&\cite {Oudichhya:2022ssc}
			&\cite {Giannuzzi:2009gh}
			&\cite {Eakins:2012jk}
			&\cite {Brown:2014ena}
			&\cite {Karliner:2014gca}
			&\cite {Mathur:2018epb}
			&\cite {Roncaglia:1995az}
			\\\hline
			&$\Xi_{bc}$&$1/2^+$ &6979&7000&6963&6955&7047&$\cdot\cdot\cdot$&6920&7037&6959&6933&6966&7040\\
			&$\Xi_{bc}^{\prime}$&$1/2^+$ &6953&6950&6933&6952&7011&6902/6906&6904&7014&6943&6914&6945&6990\\
			&$\Xi_{bc}^{\prime*}$&$3/2^+$ &6997&7020&6980&6980&7074&7030/7029&6936&7064&6985&6969&6989&7060\\
			\hline
			&States
			&$J^{P}$
			&Our work
			&\cite{Ebert:1996ec}
			&\cite{Ebert:2002ig}
			&\cite{Li:2022ywz}
			&\cite{Roberts:2007ni}
			&\cite {Oudichhya:2022ssc}
			&\cite {Giannuzzi:2009gh}
			&\cite {Tong:1999qs}
			&\cite {Brown:2014ena}
			&\cite {Mathur:2018epb}
			&\cite {Roncaglia:1995az}
			\\\hline
			&$\Omega_{bc}$&$1/2^+$ &7109&7090&7116&7055&7165&$\cdot\cdot\cdot$&7005&7110&7032&7045&7090&$\cdot\cdot\cdot$\\
			&$\Omega_{bc}^{\prime}$&$1/2^+$ &7092&7050&7088&7053&7136&7035&6994&7050&7045&6994&7060&$\cdot\cdot\cdot$\\
			&$\Omega_{bc}^{\prime*}$&$3/2^+$&7125&7110&7130&7079&7187&7149&7017&7130&7059&7056&7120&$\cdot\cdot\cdot$\\
			
		\end{tabular}
	\end{ruledtabular}
\end{table*}

Unlike singly heavy baryons, the  heavy quark subsystem is more easily excited owing to its  larger reduced mass, and then the $\rho-$mode excited states are lower than the $\lambda-$mode ones. It can be seen that our calculated mass spectra for bottom-charmed baryons  faithfully reflect this specific feature. Meanwhile, the fine structures of excited states are small and the spectra are highly degenerate, especially for the $D-$wave states. Therefore, the low-lying $P-$wave excitations are more likely to be recognized both theoretically and experimentally. Moreover, the mass spectra for bottom-charmed baryons show quite similar patterns as other heavy-light systems, such as conventional charmed or bottom  mesons, which suggests that the approximate light flavor SU(3) symmetry and heavy super-flavor symmetry are preserved well. Indeed, analogous to the doubly bottom baryons~\cite{He:2021iwx}, the bottom and charm quarks stay close to each other like a static color source, and the light quark is shared by these two heavy quarks.

Here, we can also discuss the theoretical uncertainties arising from  parameters $\alpha_{ss}$  and $\alpha_{so}$ for doubly heavy baryons.  We first vary  $\alpha_{ss}$ in the range of 1.10$\sim$1.30, and find that the discrepancies between theoretical results and experimental data for the mass splittings of singly heavy baryons can reach up to 39 MeV. Even with this large variation, the uncertainties for doubly heavy baryons are
about 7 MeV, which are small enough. The same procedure should also be done for $\alpha_{so}$, but the established excited heavy baryons relevant with this parameter are few. Here, we vary this value in a wide range of 0 $\sim$ 0.154, and  find the uncertainties for
doubly heavy baryons are about 10 MeV, which suggests that our predictions are  also stable against the parameter $\alpha_{so}$.

\subsection{Strong decays for $\lambda-$mode  $\Xi_{bc}$  and $\Omega_{bc}$ states}
The  strong decays for $\lambda-$mode $\Xi_{bc}$ and $\Omega_{bc}$ states are calculated and listed in Table~\ref{xi cc1p states} and  ~\ref{xi cb2s states}. For the two $\Xi_{bc}(1P)$ states, they have the light quark spins $j=1/2 $ and $3/2$ , and  the decay widths for $\Xi_{bc}(1/2^{-},1/2)$ and $\Xi_{bc}(3/2^{-},3/2)$ states are about 95 and 21 MeV, respectively. Owing to the limited phase space, both of them can only decay into the $\Xi_{bc}\pi$ channel. Because of relevant partial waves of decaying channels, we have found a rather broad $J^P=1/2^-$ state with $S-$wave decay and a narrow $J^P=3/2^-$ state  with $D-$wave decay, which are roughly consistent with previous work~\cite{Eakins:2012fq}. The distinctions of predicted decay widths may arise from the different phase spaces, wave functions, and phenomenological models.  
 
\begin{table}[htbp]
	\caption{\label{xi cc1p states}The predicted strong decay widths of $\Xi_{bc}(1P)$ and $\Omega_{bc}(1P)$ states in MeV. The $\cdot\cdot\cdot$ stands for the closed channel.}
	\begin{ruledtabular}
		\begin{tabular}{ccccccccc}
			&State
			&$\Xi_{bc}(\frac{1}{2}^{-},\frac{1}{2})$
			&$\Xi_{bc}(\frac{3}{2}^{-},\frac{3}{2})$
			\\\hline
			&$\Xi_{bc}$ $\pi$&94.98&20.89\\
			&Total &94.98&20.89\\
			\hline
			&State
			&$\Omega_{bc}(\frac{1}{2}^{-},\frac{1}{2})$
			&$\Omega_{bc}(\frac{3}{2}^{-},\frac{3}{2})$
			\\\hline
			&$\Xi_{bc}$ $\bar{K}$&$\cdot\cdot\cdot$&$\cdot\cdot\cdot$\\
			&Total &Narrow&Narrow\\
		\end{tabular}
	\end{ruledtabular}
\end{table}

\begin{table*}[htbp]
	\caption{\label{xi cb2s states}The predicted strong decay widths of $\Xi_{bc}(2S,1D)$ and $\Omega
		_{bc}(2S,1D)$ states in MeV.}
	\begin{ruledtabular}
		\begin{tabular}{cccccccccc}
			&State~~~~~~~~
			&$\Xi_{bc}(2S)$
			&$\Xi_{bc}(\frac{3}{2}^{+},\frac{3}{2})$
			&$\Xi_{bc}(\frac{5}{2}^{+},\frac{5}{2})$
			\\\hline
			&$\Xi_{bc}$ $\pi$~~~~~~~~&370.92&297.85&142.99\\
			&$\Xi_{bc}$ $\eta$~~~~~~~~&68.18&155.72&145.35\\
			&$\Omega_{bc}$ $\bar K$~~~~~~~~&22.28&87.85&89.66\\
			&$\Xi_{bc}(\frac{1}{2}^{-},\frac{1}{2})$ $\pi$~~~~~~~~&1.52&18.91&15.06\\
			&$\Xi_{bc}(\frac{3}{2}^{-},\frac{3}{2})$ $\pi$~~~~~~~~&2.30&27.96&32.41\\
			&Total ~~~~~~~~&465.20&588.29&425.47\\
			\hline
			&State~~~~~~~~
			&$\Omega_{bc}(2S)$
			&$\Omega_{bc}(\frac{3}{2}^{+},\frac{3}{2})$
			&$\Omega_{bc}(\frac{5}{2}^{+},\frac{5}{2})$
			\\\hline
			&$\Xi_{bc}$ $\bar{K}$~~~~~~~~&48.19&438.76&63.59\\
			&$\Omega_{bc}$ $\eta$~~~~~~~~&0.08&24.71&0.32\\
			&Total ~~~~~~~~&48.27&463.47&63.91\\
		\end{tabular}
	\end{ruledtabular}
\end{table*}

For the  two $\Omega_{bc}(1P)$ states, they lie below the $\Xi_{bc} \bar{K}$ threshold, and then the OZI-allowed strong decay is forbidden. The dominating decay channels should be $\Omega_{bc}\pi$ and $\Omega_{bc}\gamma$. This situation is quite similar to $D_{s0}^{*}(2317)$ and $ D_{s1}(2460)$ resonances, where the isospin breaking decay and radiative decay modes dominate. Actually, based on the heavy super-flavor symmetry~\cite{Savage:1990di}, the two $\Omega_{bc}(1P)$ states can be related to the charmed mesons and $\Omega_{bb}(1P)$ states~\cite{He:2021iwx}. More theoretical studies on this topic can help us to better understand the $D_{s0}(2317)$ resonance.

From the Table ~\ref{xi cb2s states}, it can be seen that most of the $2S$ and $1D$ states are broad, which can hardly be observed in experiments. For the $\Omega_{bc}(2S)$ state, the total width is relatively small,  and the dominant decay mode is $\Xi_{bc}\bar{K}$  within leading terms of the nonrelativistic transition amplitude Eq.~(\ref{2}). Compared to the $\Omega_{bc}(2S)$ state, the $\Xi_{bc}(2S)$ state is broad, which results from the larger momentum of emitted pion and phase space. If we adopt a smaller initial hadronic mass, the partial width of $\Xi_{bc}\pi$ mode and total decay width will decrease rapidly. We will discuss the relativistic corrections for these radially excited states in the following subsection. Moreover, the  narrow $\Omega_{bc}(5/2^{+},5/2)$ state mainly decays  into the  $\Xi_{bc}\bar{K}$ final states, which can be tested by future experiments.

\subsection{Strong decays for $\lambda-$mode $\Xi_{bc}^{\prime}$  and $\Omega_{bc}^{\prime}$ states}

The  strong decays for $\lambda-$mode $\Xi_{bc}^{\prime}$ and $\Omega_{bc}^{\prime}$ states are estimated and shown in Table~\ref{omega cc1p states} and ~\ref{omega cb2s states}. In the $j-j$ coupling scheme, there are five $\lambda$ mode $\Xi_{bc}^{\prime}(1P)$ states, which can be classified into two groups according to the light quark spin $j$: $j=1/2$ doublet and $j=3/2$ triplet. For the $j=1/2$ doublet, the calculated decay widths are rather broad with the current mass predictions of initial and final states, which agree with the calculations in the quark pair creation model~\cite{Eakins:2012fq}. For the $j=3/2$ triplet, the predicted decay widths are about 65, 75, and 92  MeV for the $J^{P}=1/2^{-}$, $ 3/2^{-}$, and $5/2^{-}$ states, respectively. The strong decay for $\Xi_{bc}(1/2^{-},3/2)$ is governed by $\Xi_{bc}^{\prime*}\pi$ decay mode , while  $\Xi_{bc}(3/2^{-},3/2)$ and $\Xi_{bc}(5/2^{-},3/2)$ states can decay into both $\Xi_{bc}^{\prime}\pi$ and $\Xi_{bc}^{\prime*}\pi$ channels. The broad $j=1/2$ doublet and narrow $j=3/2$ triplet are expected by the heavy quark symmetry, which arise from the enhancement or cancellation in the amplitude with different Clebsch-Gordan coefficients.     

\begin{table}[htbp]
	\caption{\label{omega cc1p states}The predicted strong decay widths of $\Xi_{bc}^{\prime}(1P)$ and $\Omega_{bc}^{\prime} (1P)$ states in MeV. The $\cdot\cdot\cdot$ stands for the closed channel, and $\times$ denotes the forbidden channel due to quantum numbers. }
	\begin{ruledtabular}
		\begin{tabular}{ccccccc}
			&State
			&$\Xi_{bc}^{\prime}(\frac{1}{2}^{-},\frac{1}{2})$
			&$\Xi_{bc}^{\prime}(\frac{3}{2}^{-},\frac{1}{2})$
			&$\Xi_{bc}^{\prime}(\frac{1}{2}^{-},\frac{3}{2})$
			&$\Xi_{bc}^{\prime}(\frac{3}{2}^{-},\frac{3}{2})$
			&$\Xi_{bc}^{\prime}(\frac{5}{2}^{-},\frac{3}{2})$
			\\\hline
			&$\Xi_{bc}^{\prime}$ $\pi$&448.45&×&×&20.21&58.43\\
			&$\Xi_{bc}^{\prime*}$ $\pi$&×&334.12&65.40&54.50&33.97\\
			&Total &448.45&334.12&65.40&74.71&92.40\\
			\hline\hline
			&State
			&$\Omega_{bc}^{\prime}(\frac{1}{2}^{-},\frac{1}{2})$
			&$\Omega_{bc}^{\prime}(\frac{3}{2}^{-},\frac{1}{2})$
			&$\Omega_{bc}^{\prime}(\frac{1}{2}^{-},\frac{3}{2})$
			&$\Omega_{bc}^{\prime}(\frac{3}{2}^{-},\frac{3}{2})$
			&$\Omega_{bc}^{\prime}(\frac{5}{2}^{-},\frac{3}{2})$
			\\\hline
			&$\Xi_{bc}^{\prime}$ $\bar{K}$&$\cdot\cdot\cdot$&×&×&0.00&0.08\\
			&$\Xi_{bc}^{\prime*}$ $\bar{K}$&×&$\cdot\cdot\cdot$&$\cdot\cdot\cdot$&$\cdot\cdot\cdot$&0.90\\
			&Total &Narrow&Narrow&Narrow&0.00&0.98\\
		\end{tabular}
	\end{ruledtabular}
\end{table}

\begin{table*}[htbp]
	\caption{\label{omega cb2s states}The predicted strong decay widths of $\Xi_{bc}^{\prime}(2S,1D)$ and $\Omega_{bc}^{\prime}(2S,1D)$ states in MeV. The $\cdot\cdot\cdot$ stands for the closed channel, and $\times$ denotes the forbidden channel due to quantum numbers. }
	\begin{ruledtabular}
		\begin{tabular}{cccccccccc}
			&State~~~~~~~~
			&$\Xi_{bc}^{\prime}(2S)$
			&$\Xi_{bc}^{\prime*}(2S)$
			&$\Xi_{bc}^{\prime}(\frac{1}{2}^{+},\frac{3}{2})$
			&$\Xi_{bc}^{\prime}(\frac{3}{2}^{+},\frac{3}{2})$
			&$\Xi_{bc}^{\prime}(\frac{5}{2}^{+},\frac{3}{2})$
			&$\Xi_{bc}^{\prime}(\frac{3}{2}^{+},\frac{5}{2})$
			&$\Xi_{bc}^{\prime}(\frac{5}{2}^{+},\frac{5}{2})$
			&$\Xi_{bc}^{\prime}(\frac{7}{2}^{+},\frac{5}{2})$
			\\\hline
			&$\Xi_{bc}^{\prime}$ $\pi$~~~~~~~~&11.11&17.49&66.05&56.84&×&×&6.72&16.21\\
			&$\Xi_{bc}^{\prime}$ $\eta$~~~~~~~~&1.57&8.52&37.13&27.10&×&×&0.68&1.71\\
			&$\Omega_{bc}^{\prime}$ $\bar K$~~~~~~~~&0.03&0.20&23.40&15.12&×&×&0.34&0.91\\
			&$\Xi_{bc}^{\prime*}$ $\pi$~~~~~~~~&78.72&51.96&7.23&33.92&76.33&30.10&21.84&12.33\\
			&$\Xi_{bc}^{\prime*}$ $\eta$~~~~~~~~&5.30&5.85&4.14&16.80&37.95&2.16&1.70&1.01\\
			&$\Omega_{bc}^{\prime*}$ $\bar K$~~~~~~~~&$\cdot\cdot\cdot$&0.00&3.02&8.09&18.40&0.94&0.75&2.93\\
			&$\Xi_{bc}^{\prime}(\frac{1}{2}^{-},\frac{1}{2})$ $\pi$~~~~~~~~&0.01&×&×&0.08&0.22&7.76&8.07&×\\
			&$\Xi_{bc}^{\prime}(\frac{3}{2}^{-},\frac{1}{2})$ $\pi$~~~~~~~~&×&0.19&0.71&0.42&0.24&1.85&5.04&9.51\\
			&$\Xi_{bc}^{\prime}(\frac{1}{2}^{-},\frac{3}{2})$ $\pi$~~~~~~~~&×&0.20&0.20&4.99&0.84&6.58&5.03&0.29\\
			&$\Xi_{bc}^{\prime}(\frac{3}{2}^{-},\frac{3}{2})$ $\pi$~~~~~~~~&0.06&0.31&13.92&40.32&7.88&6.55&6.60&4.79\\
			&$\Xi_{bc}^{\prime}(\frac{5}{2}^{-},\frac{3}{2})$ $\pi$~~~~~~~~&0.11&0.20&3.08&6.05&44.96&2.10&7.69&13.84\\
			&Total ~~~~~~~~&96.91&84.92&158.88&209.73&186.82&58.04&64.46&63.53\\
			\hline
			&State~~~~~~~~
			&$\Omega_{bc}^{\prime}(2S)$
			&$\Omega_{bc}^{\prime*}(2S)$
			&$\Omega_{bc}^{\prime}(\frac{1}{2}^{+},\frac{3}{2})$
			&$\Omega_{bc}^{\prime}(\frac{3}{2}^{+},\frac{3}{2})$
			&$\Omega_{bc}^{\prime}(\frac{5}{2}^{+},\frac{3}{2})$
			&$\Omega_{bc}^{\prime}(\frac{3}{2}^{+},\frac{5}{2})$
			&$\Omega_{bc}^{\prime}(\frac{5}{2}^{+},\frac{5}{2})$
			&$\Omega_{bc}^{\prime}(\frac{7}{2}^{+},\frac{5}{2})$
			\\\hline
			&$\Xi_{bc}^{\prime}$ $\bar{K}$~~~~~~~~&1.79&4.79&95.84&72.40&×&×&6.36&15.01\\
			&$\Omega_{bc}^{\prime}$ $\eta$~~~~~~~~&0.00&0.04&6.66&3.62&×&×&0.14&0.06\\
			&$\Xi_{bc}^{\prime*}$ $\bar{K}$~~~~~~~~&16.74&7.89&12.65&54.71&110.63&20.58&23.43&8.15\\
			&$\Omega_{bc}^{\prime*}$ $\eta$~~~~~~~~&$\cdot\cdot\cdot$&0.00&1.07&1.66&3.74&0.25&0.21&0.02\\
			&Total ~~~~~~~~&18.53&12.72&116.22&132.39&114.37&20.83&30.14&23.24\\
			
		\end{tabular}
	\end{ruledtabular}
\end{table*}

For the five $\Omega_{bc}^{\prime}(1P)$ states, the predicted masses are below or near threshold, and the total decay widths are extremely narrow. Also, according to the heavy supper-flavor symmetry, these states may have similar properties to $D_{s}(1P)$ mesons, such as the mysterious $D_{s0}^{*}(2317)$ state. These states can be hunted for through the pion and photon emissions in future experiments. For $\Omega_{bc}^{\prime}(5/2^{-},3/2)$ state, it can also be observed in the $\Xi_{bc}^{\prime}\bar{K}$ and  $\Xi_{bc}^{\prime*}\bar{K}$ invariant masses.  Moreover, the narrow $\Omega_{bc}^{\prime}(1P)$ states may be observed more easily than the ground states in the future
as well as the  singly bottom $\Omega_{b}$ family. 

For the radially excited  $  \Xi_{bc}(2S)$ and $\Omega_{bc}(2S)$ states, our calculated decay widths are relatively narrow,  and the dominant decay modes are $\Xi_{bc}^{\prime*}\pi$ and $\Xi_{bc}^{\prime*}\bar{K}$ with  the nonrelativistic transition amplitude Eq.~(\ref{2}), respectively. These relatively  narrow decays width for radially excited state are usually obtained within the nonrelativistic reduction of the axial-vector coupling between the pseudoscalar meson and  light quark. This is due to the kind of selection rule from the structure of the transition operator in the leading order of nonrelativistic expansion, and consequently due to the orthogonality of the orbital wave functions between initial and final baryons. We will continue to discuss the  relativistic corrections for Roper-like resonances in the following subsection.

For the $\Xi_{bc}^{\prime}(1D)$ and $\Omega_{bc}^{\prime}(1D)$ states, they can be divided into $j=3/2$ and $j=5/2$ triplets. the $j=3/2$ triplet are predicted to be relatively broad, while the $j=5/2$ are narrow states. It can be seen that the approximate heavy quark symmetry is preserved  well in present calculations.

\subsection{Relativistic corrections of order $1/m^{2}$ for Roper-like resonances}
The relativistic  corrections of order $1/m^{2}$ for convetional baryons are investigated in Refs.~\cite{Arifi:2021orx, Arifi:2022ntc}. The authors found that these corrections of order $1/m^{2}$ are significant for the radially excited states, that is Roper-like resonances, while the effects for  $P-$wave and $D-$wave states are small enough. In the present work, we also investigate these relativistic corrections for six radially excited bottom-charmed baryons. 

We only take into account the ground states in the final states for comparison. The results are listed in  Table~\ref{re}, and it can be seen that these relativistic corrections for Roper-like resonances are significant. This specific feature has been found in other Roper-like resonances in the literature. Owning to the lack of experimental information for bottom-charmed baryons, more theoretical and experimental efforts are needed for further exploration.

\begin{table}[htbp]
\caption{\label{xi cb2s states}The predicted strong decays into ground states for $\Xi_{bc}(2S)$,  $\Omega_{bc}(2S)$, $\Xi_{bc}^{\prime}(2S)$, and $\Omega_{bc}^{\prime}(2S)$ states in nonrelativistic transitions together with relativistic corrections (NR+RC) in MeV. The $\cdot\cdot\cdot$ stands for the closed channel. }
\begin{ruledtabular}
\begin{tabular}{cccccccccc}{\label{re}}
&State~~~~~~~~
&$\Xi_{bc}(\Gamma_{NR}+\Gamma_{RC})$
\\\hline
&$\Xi_{bc}$ $\pi$~~~~~~~~&370.92+274.00\\
&$\Xi_{bc}$ $\eta$~~~~~~~~&68.18+0.25\\
&$\Omega_{bc}$ $\bar K$~~~~~~~~&22.28+19.24\\
&Total~~~~~~~~&461.38+293.49\\
\hline
&State~~~~~~~~
&$\Omega_{bc}(\Gamma_{NR}+\Gamma_{RC})$
\\\hline
&$\Xi_{bc}$ $\bar{K}$~~~~~~~~&0.08+186.70\\
&$\Omega_{bc}$ $\eta$~~~~~~~~&48.19+0.81\\
&Total~~~~~~~~&48.27+187.51\\
\hline
&State~~~~~~~~
&$\Xi_{bc}^{\prime}(\Gamma_{NR}+\Gamma_{RC})$
&$\Xi_{bc}^{\prime*}(\Gamma_{NR}+\Gamma_{RC})$
\\\hline
&$\Xi_{bc}^{\prime}$ $\pi$~~~~~~~~&11.11+31.33&17.49+131.87\\
&$\Xi_{bc}^{\prime}$ $\eta$~~~~~~~~&1.57+0.00&8.52+0.26\\
&$\Omega_{bc}^{\prime}$ $\bar K$~~~~~~~~&0.03+0.06&0.20+12.91\\
&$\Xi_{bc}^{\prime*}$ $\pi$~~~~~~~~&78.72+121.04&51.96+120.38\\
&$\Xi_{bc}^{\prime*}$ $\eta$~~~~~~~~&5.30+0.41&5.85+17.50\\
&$\Omega_{bc}^{\prime*}$ $\bar K$~~~~~~~~&$\cdot\cdot\cdot$&0.00+0.00\\
&Total~~~~~~~~&96.73+152.84&84.02+282.92\\
\hline
&State~~~~~~~~
&$\Omega_{bc}^{\prime}(\Gamma_{NR}+\Gamma_{RC})$
&$\Omega_{bc}^{\prime*}(\Gamma_{NR}+\Gamma_{RC})$
\\\hline
&$\Xi_{bc}^{\prime}$ $\bar K$~~~~~~~~&1.79+31.74&4.79+154.91\\
&$\Omega_{bc}^{\prime}$ $\eta$~~~~~~~~&0.00+0.31&0.04+2.93\\
&$\Xi_{bc}^{\prime*}$ $\bar K$~~~~~~~~&16.74+139.40&7.89+113.66\\
&$\Omega_{bc}^{\prime*}$ $\eta$~~~~~~~~&$\cdot\cdot\cdot$&0.00+0.25\\
&Total~~~~~~~~&18.53+171.45&12.72+271.75\\
\end{tabular}
\end{ruledtabular}
\end{table}

 \subsection{Mixing}

 In our calculation, we adopt the $j-j$ coupling scheme and the basis in the heavy quark limit to study mass spectra and strong decays for bottom-charmed baryons. Due to the finite mass of heavy quark subsystem, the physical observed resonances may correspond to the superposition of theoretical states in the quark model. For instance, the mixing scheme for $\lambda-$mode $\Xi_{bc}^{\prime}(1P)$ and $\Omega_{bc}^{\prime}(1P)$ states can be formulated as
 \begin{equation}
 \left(\begin{array}{cc}
\mid 1 P & \left.1 / 2^{-}\right\rangle_{1} \\
\mid 1 P & \left.1 / 2^{-}\right\rangle_{2}
\end{array}\right)=\left(\begin{array}{cc}
\cos \theta & \sin \theta \\
-\sin \theta & \cos \theta
\end{array}\right)\left(\begin{array}{l}
\left|1 / 2^{-}, j=1/2\right\rangle \\
\left|1 / 2^{-}, j=3/2\right\rangle
\end{array}\right),
 \end{equation}
 \begin{equation}
 \left(\begin{array}{cc}
\mid 1 P & \left.3 / 2^{-}\right\rangle_{1} \\
\mid 1 P & \left.3 / 2^{-}\right\rangle_{2}
\end{array}\right)=\left(\begin{array}{cc}
\cos \theta & \sin \theta \\
-\sin \theta & \cos \theta
\end{array}\right)\left(\begin{array}{l}
\left|3 / 2^{-}, j=1/2\right\rangle \\
\left|3 / 2^{-}, j=3/2\right\rangle
\end{array}\right),
 \end{equation}
where $\theta$ is the mixing angle. Also, the $D-$wave excited states with the same spin-parity can mix with each other.

In the heavy quark limit, the mixing angle should be zero. Actually,  the heavy quark subsystem including bottom and charm quarks are relatively heavy, and  heavy quark symmetry should be  approximately preserved. Then, the above mixing angles are expected to be  small enough. We calculate the mixture for $\lambda-$mode $\Xi_{bc}^{\prime}(1P)$ states as an illustration and list them in Table~\ref{mix}. It can be found that the mixing angles are tiny, and the mixing effects for spectroscopy can be neglected in bottom-charmed sector. Moreover, there may also  exist other mixing scheme, which are believed to be even smaller in these bottom-charmed baryons.

\begin{table}[htp]
	\begin{center}
		\caption{\label{mix}The superposition of $\Xi_{bc}^{\prime}(1P)$ states with $J^{P}=1/2^{-} $ and $3/2^{-}$. The mixing angles for $J^{P}=1/2^{-} $ and $3/2^{-}$ states are  and respectively.}
		\begin{tabular*}{8.5cm}{@{\extracolsep{\fill}}p{1.5cm}<{\centering}p{1.9cm}<{\centering}p{1.9cm}<{\centering}p{1.8cm}<{\centering}p{1.7cm}<{\centering}}
			\hline\hline
			State   & $\langle H\rangle$ (MeV) & Mass (MeV)  & Eigenvector\\\hline
			$|1 P1/ 2^{-}\rangle_{1}$ & \multirow{2}{*}{$\begin{pmatrix}7396.33&-2.10\\ -2.10 & 7382.40\end{pmatrix}$}
			& \multirow{2}{*}{$\begin{bmatrix}7396.64 \\7382.09 \end{bmatrix}$}  & \multirow{2}{*}{$\begin{bmatrix}(0.99, -0.15)\\(0.15, 0.99) \end{bmatrix}$}\\
			$|1 P1/ 2^{-}\rangle_{2}$\\
			$|1 P3/ 2^{-}\rangle_{1}$ & \multirow{2}{*}{$\begin{pmatrix}7404.08&-2.85 \\ -2.85 & 7386.83 \end{pmatrix}$}
			& \multirow{2}{*}{$\begin{bmatrix}7404.54 \\7386.37 \end{bmatrix}$}  & \multirow{2}{*}{$\begin{bmatrix}(0.99, -0.16)\\(0.16, 0.99) \end{bmatrix}$}\\
			$|1 P3/ 2^{-}\rangle_{2}$\\
			\hline\hline
		\end{tabular*}
	\end{center}
\end{table}

\subsection{Low-lying $\rho-$mode and $\rho-\lambda$ hybrid states}
\begin{figure*}[! htpb]
    \centering
\includegraphics[scale=0.55]{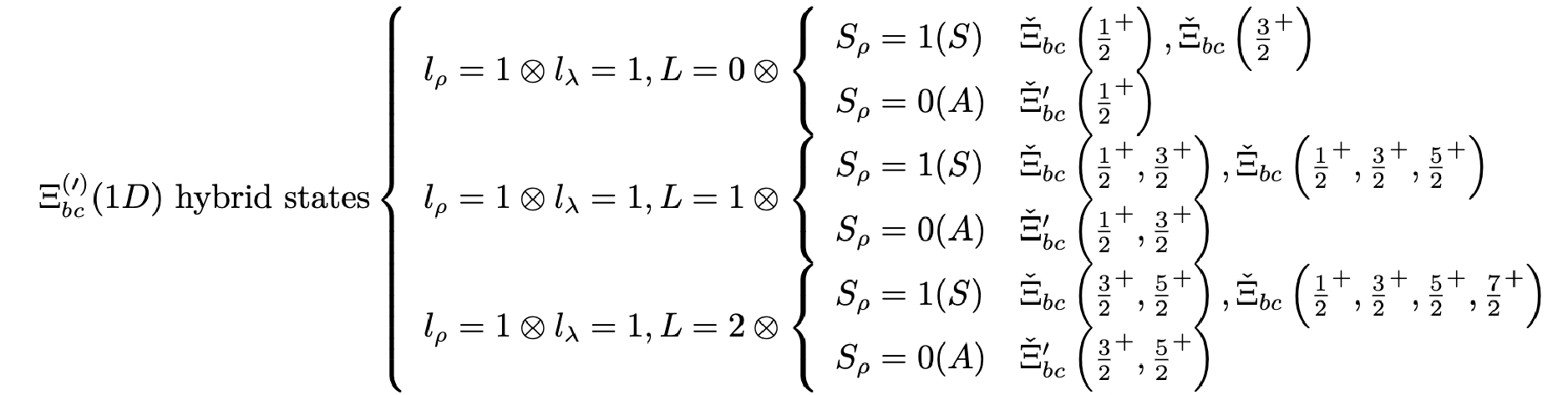}
    \caption{\label{coupling}  The $\rho-\lambda$ hybrid states for $\Xi_{bc}^{(\prime)}(1D)$  states. Here, $l_{\rho}=l_{\lambda}=1$, and then $L$ equals to 0, 1, and 2. The capital $A$ represents anti-symmetric spin wave function, and capital $S$ stands for symmetric spin wave function.}
\end{figure*}

In addition to the presence of $\lambda-$mode excitations, the bottom-charmed baryons also have low-lying $\rho-$mode and $\rho-\lambda$ hybrid states. For example,  when $l_{\rho}=l_{\lambda}=1$, there exists thirteen $\check{\Xi}_{bc}$ and five $\check{\Xi}_{bc}^{\prime}$ states, which can be seen in Figure~\ref{coupling}.

\begin{figure*}
     \centering
     \includegraphics{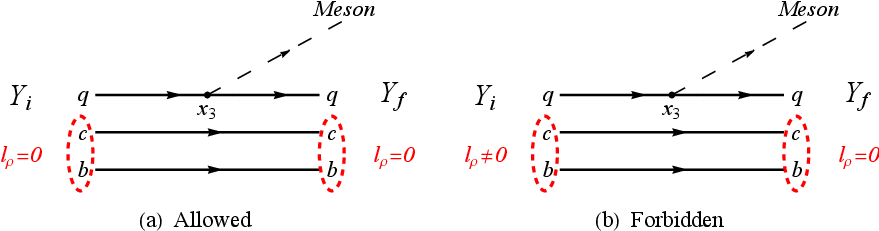}
     \caption{Schematic diagrams for strong decays of doubly heavy baryons within different exited modes.}
     \label{aorf}
 \end{figure*}

For these low-lying states, the light meson emissions are supposed to be dominating. However, 
 under the spectator assumption for the two heavy quarks, the orbital wave functions of heavy quark subsystems between initial $\rho-$mode or $\rho-\lambda$ hybrid states and final ground states are orthogonal, which results in the vanishing amplitudes and strong decay widths. More explicitly, the light meson emission occurs for $\lambda-$mode excitations, and is irrelevant to the $\rho-$mode variables.  Hence, the orthogonality of the different $\rho-$mode wave functions leads to the vanishing matrix element, which is shown in Figure \ref{aorf}.   That is to say, our calculated decay modes  preserve the heavy diquark symmetry automatically, where the heavy quark subsystems with different quantum numbers cannot transit into each other.  Thus, these states should be extremely narrow and the weak and  radiative decays  may become dominating, which  
 can provide good opportunities to be searched by future experiments. 

\section{SUMMARY}{\label{SUMMARY}}
In this work, we have studied the low-lying mass spectra for bottom-charmed baryons in a nonrelativistic quark model by solving the three-body  Schr\"odinger equation. With the obtained realistic wave functions, we get the root mean square radius and study the strong decays of bottom-charmed baryons. The lowest $\Xi_{bc}$, $\Xi_{bc}^\prime$, $\Omega_{bc}$, and $\Omega_{bc}^\prime$ states are predicted to be about 6979, 6953, 7109, and 7092 MeV, respectively. Our results indicate that some of $\lambda-$mode $\Xi_{bc}(1P)$, $\Xi_{bc}^{\prime}(1P)$, $\Omega_{bc}(1P)$, $\Omega_{bc}^{\prime}(1P)$ states are relatively narrow, which have good potentials to be observed by future  experiments. Also, the strong decays of the low-lying $\rho-$mode and $\rho-\lambda$ hybrid states are highly suppressed and  can be hunted for in the electroweak processes. 

Given the heavy quark symmetry, the heavy quark subsystem in bottom-charmed baryons play a role as a heavy antiquark, and then the heavy super-flavor symmetry emerges. Indeed, our results about the mass spectra and strong decays for bottom-charmed baryons support this claim. With the development of the large-scale accelerator facilities, we expect that more theoretical and experimental efforts are involved to search for more doubly heavy baryons and better understand the heavy quark symmetry.

\section*{ACKNOWLEDGMENTS}
This work is supported by the National Natural Science
Foundation of China under Grants No. 11705056, the Natural Science Foundation of Hunan Province under Grant No. 2023JJ40421, the Key Project of Hunan Provincial Education Department under Grant No. 21A0039,  and the State
Scholarship Fund of China Scholarship Council under Grant
No. 202006725011. A. H. is supported by the Grants-in-Aid for Scientific Research (Grant Numbers 21H04478(A))
and the one on Innovative Areas (No. 18H05407).

\end{document}